\documentclass[12pt,reqno, a4paper]{amsart}
\usepackage{graphicx}
\usepackage{fancyhdr}
\usepackage{enumerate}
\usepackage{amsmath}
\usepackage{setspace}
\usepackage{capt-of}
\usepackage[font=sf]{caption}

\usepackage{amsthm}
\usepackage{mathabx}
\usepackage{amssymb}
\usepackage{relsize}
\usepackage{pdfsync}
\usepackage[colorlinks=true,linkcolor=blue,citecolor=magenta]{hyperref}

\usepackage[notcite,notref,color]{showkeys}

\usepackage[normalem]{ulem}

\usepackage{xcolor}

\usepackage{calligra}

\usepackage{tikz}

\usepackage{changebar}
\usepackage{pdfsync}
\usetikzlibrary{decorations.pathmorphing,patterns}
\usetikzlibrary{calc,arrows}

\usepackage{pgf}

\usetikzlibrary{automata}
\usetikzlibrary{positioning}

\tikzset{
    state/.style={
           rectangle,
           rounded corners,
           draw=black, very thick,
           minimum height=2em,
           inner sep=2pt,
           text centered,
           },
}

\usepackage{MnSymbol}

\usepackage{xcolor}

\usepackage{changebar}
\usepackage{pdfsync}

\usepackage[normalem]{ulem}

\makeindex
\newtheorem{theorem}{Theorem}[section]

\newtheorem{remark}[theorem]{Remark}

\newcommand{\cal}{\mathcal}

\newcommand\bbR{{\mathbb R}}
\newcommand\bbZ{{\mathbb Z}}

\newcommand\al{\alpha}

\newcommand \ga{\gamma}
\newcommand \om{\omega}

\renewcommand{\ge}{\geqslant}
\renewcommand{\le}{\leqslant}

\renewcommand{\tilde}{\widetilde}
\renewcommand{\bar}{\overline}
\numberwithin{equation}{section}

\newcommand{\tk}[1]{{\color{blue}#1}}

\begin{document}

\title{Periodically Driven anharmonic chain: Convergent Power Series and Numerics }

\author{Pedro L. Garrido}
 \address{Pedro L. Garrido\\Universidad de Granada\\Granada Spain} 
\email{{\tt garrido@onsager.ugr.es}}

 \author{Tomasz Komorowski}
 \address{Tomasz Komorowski, Institute of Mathematics,
   Polish Academy Of Sciences, Warsaw, Poland.} 
\email{{\tt tkomorowski@impan.pl}}

\author{Joel L. Lebowitz}
\address{Joel L. Lebowitz, Departments of Mathematics and Physics,  Rutgers University}
\email{\tt lebowitz@math.rutgers.edu}

 \author{Stefano Olla}
 \address{Stefano Olla, CEREMADE,
   Universit\'e Paris-Dauphine, PSL Research University \\
 \and Institut Universitaire de France\\ \and
GSSI, L'Aquila}
  \email{\tt olla@ceremade.dauphine.fr}

\date{\today {\bf File: {\jobname}.tex.}}

\begin{abstract}

 We investigate the long time behavior of a pinned chain of $2N+1$
 oscillators, indexed by   $x \in\{-N,\ldots, N\}$.  The system is subjected to an external driving force on the particle at $x=0$, of period $\theta=2\pi/\omega$, and to frictional
 damping  $\gamma>0$ at both endpoints $x=-N$ and $N$.
 The oscillators interact with a pinned and nearest neighbor harmonic 
 plus  anharmonic potentials of the form
 $\frac{\om_0^2 q_x^2}{2}+\frac12 (q_{x}-q_{x-1})^2 +\nu\left[V(q_x)+U(q_x-q_{x-1}) \right]$, with $V''$ and $U''$
 bounded and $\nu\in\bbR$.
 We recall the recently proven convergence and the global stability of
 a perturbation series in powers of $\nu$ for
 $|\nu| < \nu_0$, yielding the long time periodic state of the
 system.
 Here $\nu_0$ depends only on the supremum norms of $V''$ and $U''$
 and the distance of the set of non-negative integer multiplicities of
 $\om$ from the interval $[\om_0,\sqrt{\om_0^2+4}]$ - the spectrum of
 the infinite harmonic chain for $\nu=0$. We describe also some
numerical studies of this system going beyond our   rigorous results.
 
\end{abstract}

  \thanks{} 

\maketitle


\section{Introduction}

In a recent work \cite{our} we derived new results for the time evolution
of a finite pinned anharmonic chain of $2N+1$ oscillators,
indexed by $x \in\{-N,\ldots, N\}$, $N=0,1,\ldots$ subjected to an external
driving force  ${\cal F}(t/\theta)$ of
period $\theta>0$,   acting
on the oscillator at $x=0$. In this note we first describe briefly the rigorous results of \cite{our} and then present new numerical results
for values of the parameters not covered there. 

The Hamiltonian of the chain is given by:
\begin{equation}
\label{Hn}
\mathcal{H}_N (\mathbf q, \mathbf p;\nu):=
\sum_{x=-N}^N \left[ \frac{p_x^2}2 +
\frac12 (q_{x}-q_{x-1})^2 +\frac{\om_0^2 q_x^2}{2}+\nu\Big( V(q_x)+  U(q_x-q_{x-1})\Big)\right],
\end{equation}
where $(\mathbf q, \mathbf p)=\big(q_x,p_x\big)_{x\in \bbZ_N}$ is the configuration of the positions and momenta of the oscillators and $q_{-N-1}=q_{-N}$.

 The  microscopic dynamics of
the process  
 is  given   by the forced Hamiltonian
system with friction on both endpoints 
\begin{equation} 
\label{eq:HZ100}
\begin{aligned}
   \ddot    q_x(t;\nu) &=  \Delta q_x(t;\nu)-\om_0^2 q_x(t;\nu)  -\ga
   \dot q_x(t;\nu)\delta_{x,-N}-\ga
   \dot q_x(t;\nu)\delta_{x,N}\\
  &
 -\nu\Big( V'(q_x(t;\nu)) -\nabla U'\big(q_x(t;\nu)-q_{x-1}(t;\nu)\big) \Big)+ {\cal F}(t/\theta)\delta_{x,0} ,\quad
 x\in \bbZ_N.
\end{aligned} \end{equation}
where $\ga>0$ is the friction coefficient and $p_x(t;\nu)=\dot q_x(t;\nu)$. 
We let the force be of the form
\begin{equation}
  \label{forcing-g}
    {\cal F}(t/\theta)=\sum_{m\in\bbZ}\tilde{\rm F}_me^{im\om t},\quad
  \end{equation}
where {$ \om=2\pi/\theta$, $\tilde{\rm F}_{0}=0$, $\tilde{\rm F}_{-m}=\tilde{\rm F}_{m}^\star$ and $ \sum_m |\tilde{\rm F}_m|^2 < +\infty$}.

 The Neumann laplacian $\Delta $ and the discrete gradient are defined as
\begin{equation}
  \label{nmn}
  \Delta f_x=f_{x+1}+f_{x-1}-2f_x,\quad \nabla
  f_x=f_{x+1}-f_x,\quad  x\in[-N,N],
\end{equation}
    {with
      the boundary condition  $f_{N+1}=f_N,\quad f_{-N-1}=f_{-N}$.

We consider the case where the
non-quadratic parts of the pinning and interacting potentials $U$ and
$V$ have bounded second derivatives
\footnote{This is more restrictive than necessary, see \cite{our}, for the proofs of our claims.}.
Examples of such potentials are furnished
by a modified FPUT potential \cite{our}
\begin{equation}
 U(r)= \frac{r^{2n}}{1+\alpha r^{2n}}\quad\mbox{for some } \,\al>0\label{pot1}
\end{equation}
or by a pinning potential \cite{Geniet}
\begin{equation}
 V(q)=  (\sin q)^{2n}\quad,\quad V(q)=a(1+q^2)^{1/2}\label{pot2}
\end{equation}
where $n$ is a positive integer and $a>0$. 

\section{Summary of the results from \cite{our}}
\label{sec2}

\subsection{The existence and stability of a periodic solution to
  (\ref{eq:HZ100})}

We proved that when all integer multiplicities of
$\omega=\frac{2\pi}{\theta}$ are outside the interval containing the
spectrum of the infinite harmonic chain, $\mathcal{I}= [\omega_0, \sqrt{\omega_0^2 +4}]$,
the unique long time state of the system is given by a
{perturbative expansion in} $\nu$, see the detailed description in
Section \ref{sec2.2} below, which 
converges for $\vert\nu\vert<\nu_0$,
where 
\begin{equation}
  \label{eq:14}
  \begin{split}
    \nu_0=\frac{\delta_*}{{\frak V}}, \qquad  \delta_* &=
    \inf\{|(m\omega)^2 - w^2| : m\in \mathbb Z, w\in\mathcal{I}= [\omega_0, \sqrt{\omega_0^2 +4}] \}\\
  {\frak V} &= \sup_q [\vert V''(q)\vert+3\vert U''(q)\vert].
\end{split}
\end{equation}

If $\vert\nu\vert<\nu_0$, then for all $N$, $\cal{F}$, $\gamma>0$  and 
any initial configuration of the chain at time $s$ , allowing $s\rightarrow -\infty$, the configuration of the chain at any time $t$
converges to  $\{q_{x,p}(t;\nu),\dot q_{x,p}(t;\nu)\}$ where $q_{x,p}(t;\nu)$ is the unique $\theta$-periodic solution of  (\ref{eq:HZ100})  achieved at large times.
More precisely, if $\{{\bf q}(t,s;\nu),\dot{\bf q}(t,s;\nu)\}$ is the solution of
(\ref{eq:HZ100}) with initial conditions ${\bf q}(s,s;\nu)={\bf q}$, $\dot{\bf q}(s,s;\nu)=\dot{\bf q}$, then

\begin{equation}
  \label{eq:13}
  \lim_{s\to-\infty} \sum_{x=-N}^N \left[(q_x\big(t,s;\nu \big)-q_{x,p}(t;\nu))^2 +
    (\dot q_x\big(t,s;\nu \big)- \dot q_{x,p}(t;\nu))^2\right] = 0. 
\end{equation}

We also show that for some $0<\tilde \nu\le \nu_0$ and all $|\nu|<\tilde \nu$, there exists $A,\rho>0$ uniform in $N$ such that
       \begin{equation}
         \label{091410-24}
         \begin{split}
            &        |q_{x,p}(t;\nu)|\le A\exp\left\{-\rho|x|\right\},\quad
            t\in[0,\theta], \quad\mbox{and}\\
            &
           \int_0^{\theta}\dot q^2_{x,p}(t;\nu)d  t\le A\exp\left\{-\rho|x|\right\}
\end{split}
\end{equation}

The exponential decay of $\dot q^2_{x,p}(t;\nu)$ with $x$ implies
that the average work performed by the external force over the period
\begin{equation}
  \label{013105-23}
  {\cal W}(\nu)=\frac{1}{\theta}\int_0^\theta {\cal F}(t/\theta) \dot q_{0,{\rm p}}(t;\nu) d  t
  =\frac{\ga}{\theta}\Big( \int_0^{\theta}\dot q_{-N, {\rm p}}^2(s)d  s
          +\int_0^{\theta}\dot q_{N, {\rm p}}^2(s)d  s\Big)>0
  \end{equation}
 goes to zero for $\vert\nu\vert<\tilde\nu<\nu_0$ when
$N\rightarrow\infty$.
 
It can be seen from the above that the convergence of a
perturbative series expansion depends on supremum norm of the second derivatives of the
 anharmonic potential being bounded, and, as far as we
know, it has not been proven for any other non-equilibrium
system. Numerical simulations shown in the present paper indicate 
that the properties of the system for $\vert\nu\vert >\nu_0$ depend  on the sign
of $\nu$ but this domain of the parameter is beyond
our present mathematical abilities.

\subsection{Perturbative scheme}
\label{sec2.2}
For the harmonic case, $\nu = 0$, we can obtain explicitly  the unique  $\theta$-periodic solution which the system
approaches after long times of order $N^3/\gamma$, see \cite{menegaki}. 
To represent a $\theta$-periodic solution  for an anharmonic chain \eqref{eq:HZ100},
  consider a perturbative series
\begin{equation}
  \label{010401-24z}
 q_{x,{\rm p}}(t;\nu)=\sum_{\ell=0}^{+\infty}q^{(\ell)}_{x,{\rm p}}(t;\nu)\nu^\ell.
\end{equation}
Its 0-th order term is given by the $\theta$-periodic solution
of the harmonic chain.
To obtain \eqref{010401-24z} we construct a sequence $\Big(Q_{x}^{(L)}(t;\nu)\Big)_{
x\in \bbZ_N}$, $L=0,1,\ldots$
of $\theta$-periodic functions that satisfy
\begin{equation} 
\label{012805-24L}
\begin{split}
  \ddot    Q_{x}^{(L)} (t;\nu)= &
  \big(\Delta  -\om_0^2\big) Q_{x}^{(L)} (t;\nu) -\ga(\delta_{x,-N}+
   \delta_{x,N} )\dot Q_{x}^{(L)}(t;\nu)
   \\
   &
   -\nu
   W_x\big(Q_{x}^{(L-1)}(t;\nu)\big)  + {\cal F}(t/\theta)\delta_{x,0} ,\quad
  x\in \bbZ_N,
   \end{split}
 \end{equation}
 {where
 \begin{equation}
  \label{W}
  W_x(f):=V'(f_x)-\nabla U'(f_x-f_{x-1}), \qquad W_x\big(Q^{(-1)}(t;\nu)\big)\equiv 0.
\end{equation}
}
{By convention $Q^{(0)}(t;\nu) :=q^{(0)}_{x,{\rm p}}(t)$ is the $\theta$-periodic solution for the harmonic case.}
For $L\ge 1$ we define
 \begin{equation}
   \label{eq:3}
   q^{(L)}_{x,{\rm p}}(t;\nu) := \nu^{-L} \left(Q_{x}^{(L)}(t;\nu) - Q_{x}^{(L-1)}(t;\nu)\right)
 \end{equation}
i.e.

\begin{equation}
  \label{010905-24}
  Q_{x}^{(L)}(t;\nu)=\sum_{\ell=0}^{L}q^{(\ell)}_{x,{\rm p}}(t;\nu)\nu^\ell.
\end{equation}

Then, by \eqref{012805-24L}, we conclude that 
$q_{x }^{(L)}(t;\nu)$, $x\in\bbZ_N$ is a $\theta$-periodic solution of 
\begin{equation} 
\label{eq:flip-lZ}
\begin{aligned}
     \ddot    q _{x,{\rm p}}^{(L)}(t;\nu) &=  \Delta_x
    q _{x,{\rm p}}^{(L)}(t;\nu)-\om_0^2  q _{x,{\rm p}}^{(L)}(t;\nu)\\
   &
   -\ga\dot
   q _{x,{\rm p}}^{(L)}(t;\nu)\delta_{-N,x}-\ga\dot
     q _{x,{\rm p}}^{(L)}(t;\nu)\delta_{N,x}-   v_{x,L-1}(t)  ,\quad
  x\in \bbZ_N,
\end{aligned} \end{equation}
where
\begin{equation}
  \label{vxL}
  \begin{split}
    &v_{x,L-1}(t)=\frac{1}{\nu^{L-1}}\Big[W_x\big(Q^{(L-1)}(t)\big)
    -W_x\big(Q^{(L-2)}(t)\big)\Big].
\end{split}
\end{equation}
From
       \eqref{eq:flip-lZ} it follows that $ q _{x,{\rm
           p}}^{(L)}(t;\nu)$ are smooth functions of $\nu$ and $t$.

     Note that
$v_{x,0}(t)= W_x\big(q_{{\rm p}}^{(0)}(t) \big)$.
In addition,  $v_{L-1}(t,\nu)$ is bounded by
       $$
     \frak{V} |Q^{(L-1)}(t) - Q^{(L-2)}(t)|.
       $$
       with $\frak{V}$ is defined in eq.\eqref{eq:14}.  
        This is where the boundedness of the second derivative of the potential comes as a crucial property.
       Using the equations for the time harmonics of $q_{x,{\rm p}}^{(L)}$ and $v_{x,L}$ we
       obtain the bound, see \cite{our} for the derivation,
      \begin{equation}
        \label{eq:9}
        \|\!|v_{L-1}(\cdot,\nu)\|\!| \le \frak V \|\!| q^{(L-1)}_{\rm p} (\cdot,\nu)\|\!|.
      \end{equation}
      
   where $\frak V $ is defined in (\ref{eq:14}) and
    \begin{equation}
     \|\!| f \|\!|^2=\frac{1}{\theta}\int_0^\theta \Big(\sum_{x=-N}^N \vert f_x(t) \vert^2\Big)dt
    \end{equation}

   Thus, we have obtained
  \begin{align}
  \label{082302-24}
  \|\!|q^{(\ell)}_{\rm p}(\cdot;\nu)\|\!|\le
    \frac{1}{\nu_0}\|\!|q^{(\ell-1)}_{\rm p}(\cdot;\nu)\|\!|,\quad \nu\in\bbR,\, 
    \ell = 1,2 \dots .  
  \end{align}
 Consequently for $|\nu|<\nu_0$
    the sums  $Q_{x,{\rm p}}^{(L)}(t;\nu)$, given by \eqref{010905-24}  converge
    in the $\|\!|\cdot\|\!|$-norm, as $L\to+\infty$.
    The convergence is uniform in $N$ and $\gamma$. Moreover
   \begin{equation}
     \label{112302-24}
     \begin{split}
     &     \|\!|\sum_{\ell= L}^{+\infty}q^{(\ell)}_{\rm p} (\cdot;\nu)\nu^\ell\|\!|
  \le \frac{|\nu/\nu_0|^L \|\!| q^{(0)}_{\rm p}(\cdot;\nu)\|\!|}{1-|\nu/\nu_0|},\quad
  L =1,2,\ldots,\,|\nu|\le \nu_0 .
\end{split}
\end{equation}
As a  result the series 
\begin{equation}
  \label{010401-24z1}
 q_{x,{\rm p}}(t;\nu)=\sum_{\ell=0}^{+\infty}q^{(\ell)}_{x,{\rm p}}(t;\nu)\nu^\ell
\end{equation}
converges for $|\nu|< \nu_0$  and 
 defines a $\theta$-periodic solution of \eqref{eq:HZ100}. In fact it
 is a
 unique    $\theta$-periodic solution for this range of anharmonicity
parameter $\nu$. We note that $\nu_0$ is a lower bound on the radius of
 convergence for any fixed $\cal{F}$, $N$ and $\gamma>0$.

 \begin{remark}
   We note here that a $\theta$-periodic solution   exists for any value
  of $\nu$. This can be seen
by an abstract compactness argument, see \cite{our} for details.
However,  as it was pointed out there, the uniqueness  does not hold in general. In fact, there are
situations where the series \eqref{010401-24z} converges for 
$\vert\nu\vert>\nu_0$ but the respective periodic solution is not
unique, nor even locally stable, see \cite{our} and Section \ref{sec3} below.
 \end{remark}

 \begin{remark}
When the periodic components of the force
$\tilde{\rm F}_m$ vanish  for all even $m$,
e.g. $\mathcal{F}(t/\theta)=F\cos(\omega t)$ and the potentials $V$
and $U$ are even functions of their argument, then  the series
\eqref{010401-24z}  converges for $|\nu|< \bar\nu_0 = \frac{\bar\delta_*}{ {\frak V}}$, where
    \[
    \bar\delta_*:=\inf\Big[\big|\big((2m-1)\om\big)^2-w^2\big|
    :\,m\in\bbZ,\,w\in {\cal I}\Big] > \delta_*.
  \]
  However in this case $q_{\rm p}(t;\nu)$ defined by \eqref{010401-24z}  need not be stable.
\end{remark}

 \begin{remark}
   Note that, the condition $|\nu|<\nu_0$ implies in particular
    that the derivative of the pinning potential, 
    $\frac{d}{dq}[\om_0^2q^2/2+\nu V(q)]$, is non-decreasing. Therefore, the
    pinning part of the potential is convex.
\end{remark}
 \begin{remark}
  It also follows from the above analysis that for   $|\nu|<\nu_0$
there exists a constant $C>0$ independent of $N$ and $\ga>0$ such
 that 
 \begin{equation}
   \label{060810-24}
  \frac{1}{\theta}\int_0^\theta{\cal H}_N\big({\bf  q}_{{\rm
      p}}(t;\nu), {\bf  p}_{{\rm p}}(t;\nu)\big)d  t\le C.
   \end{equation}
\end{remark}

\section{The one oscillator case}
\label{sec3}

We consider here the case of a single damped anharmonic
oscillator with a periodic forcing. This is of its own interest
\cite{one} and explicitly shows the dependence of the radius of
convergence of a power series \eqref{010401-24z} on the
parameters. In the present case the spectral  interval $\mathcal{I}$
is just $\omega_0$. The equation of motion is:
\begin{equation}
\ddot q=-\omega_0^2 q+2\gamma\dot q -\nu V'(q)+F\cos(\omega t) \label{one}
\end{equation}
When $\gamma>0$ and $\nu=0$, the long time state of the system given in \eqref{eq:13} is:
\begin{equation}
q_p(t,\nu=0)=\frac{F}{(\omega_0^2-\omega^2)^2+4\gamma^2\omega^2}\left[(\omega_0^2-\omega^2)\cos(\omega t)+2\gamma\omega\sin(\omega t) \right]\label{qnu0}
\end{equation}
for any initial condition. For $\nu\neq 0$, the results above shows that there exists a unique globally stable periodic solution when 
\begin{equation}
\vert\nu\vert<\nu_0=\delta_*/ \|V''\|_\infty
\end{equation}
 with  $\|\cdot\|_\infty$  the supremum norm and 
\begin{equation}
\delta_*=\inf_{m\in \mathbb Z}\vert\omega_0^2-(m\omega)^2\vert .
\end{equation}

As noted before, when restricted to the case $V(q)=V(-q)$ the power series \eqref{010401-24z} converges for
\begin{equation}
\vert\nu\vert<\bar\nu_0=\bar\delta_*/\|V''\|_\infty
\end{equation}
 with 
\begin{equation}
\bar\delta_*=\inf_{m\in \mathbb Z}\sqrt{(\omega_0^2-((2m+1)\omega)^2)^2+ (2\gamma (2m+1)\omega)^2} \geq \delta_*.
\end{equation}

We can however be sure that it is stable only when $\vert\nu\vert<\nu_0<\bar\nu_0$.

\subsection{Numerical solution for $\nu\neq 0$}

We solve numerically eq.(\ref{one}) when $V(q)=-\cos q$,  $\omega_0=1$, $\gamma=1/2$ and $F=1.75$ and $\omega=0.8$. 
The initial conditions chosen are: $q(0,\nu)=q_p(0,\nu=0)$ and $\dot q(0,\nu)=\dot q_p(0,\nu=0)$. Observe that \tk{in} this case $ \|V''\|_\infty=1$, $\bar\nu_0=0.8772..$ and  $\nu_0=0.1055...$. 

We show in Figure \ref{conff} how the values $Q^{(L)}$ of the
perturbation scheme \eqref{012805-24L} converge to the numerical
solution  for $\nu=0.8$. Moreover, we compute the distance of
$Q^{(L)}$ to the numerical periodic solution $q$ starting at
$t_0=300$  (where $q(t)$ appears to attain  the
  stationary state) during one period
$\theta$:
\begin{equation}
s(L)=\left[\frac{1}{\theta}\int_{0}^{1}d\tau \left(Q^{(L)}(t_0+\tau\theta)-q(t_0+\tau\theta)\right)^2\right]^{1/2}.\label{dist}
\end{equation}
The distance $s(L)$ decay fast with $L$ as expected. 

\begin{figure} [htbp]
 \includegraphics[width=7.1cm]{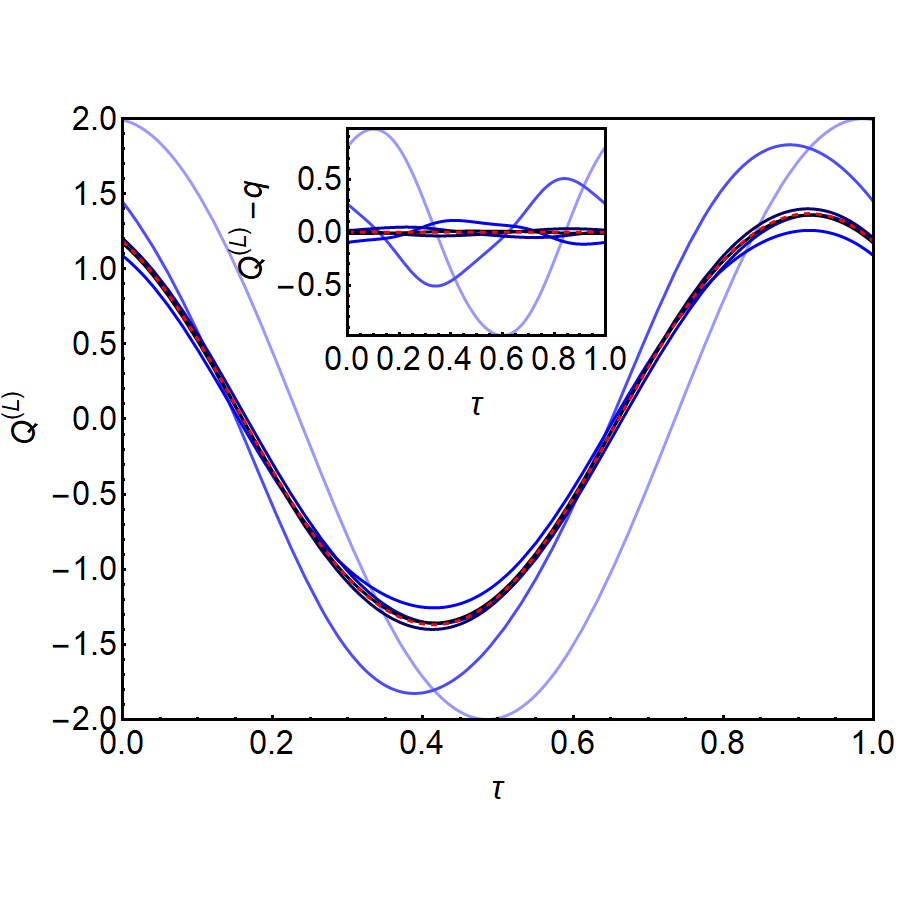} 
 \includegraphics[width=7.1cm]{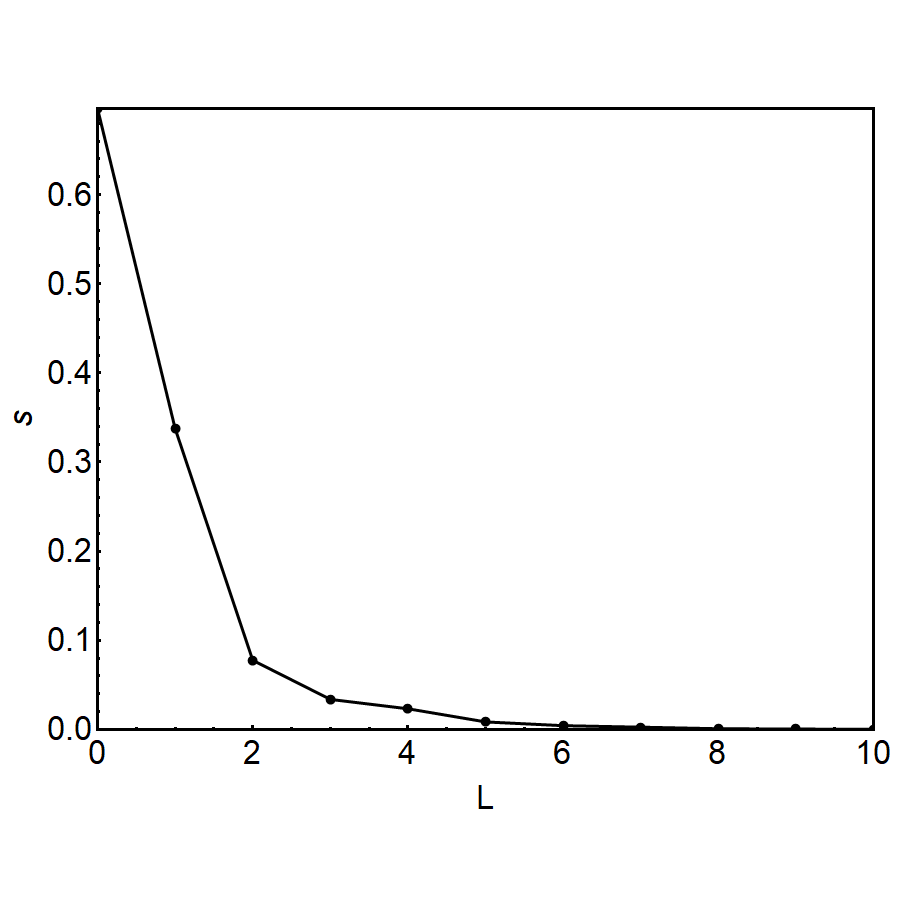}  
   \caption {Left: Numerical solution of the perturbation scheme, $Q^{(L)}$, from the differential equation \eqref{012805-24L} for $L=0, 1, 2, 3, 4$ and $5$ (from light blue to dark blue respectively) at the stationary state during one forcing period: $\tau=(t-t_0)/\theta$ with $t_0=300$. Red dashed curve is the numerical solution of the original differential equation \eqref{one}. The inset shows the difference: $Q^{(L)}(t)-q(t)$ at the stationary state during one forcing period. Right: The average distance $s(L)$, \eqref{dist},  vs $L$. The value of the parameters are discussed in the main text.} \label{conff} 
\end{figure}

We note that the value used in the computations, $\nu=0.8$, is smaller than the theoretical bound that guaranties convergence of the perturbation scheme, $\bar\nu_0$, but is larger than the bound found for the global stability, $\nu_0$. In fact, we have numerically observed a $\theta$-periodic stable solution (from $10$ different initial conditions) from  $\nu\simeq-1.6$ to $\nu$ larger than $10$. Clearly the theoretical bounds are not optimal. 
 
 Finally, out of the $\nu$-region where only the $\theta$-periodic solution exists, we find solutions with different typology. We show in Figure \ref{conf} the phase space $(q(t),\dot q(t))$ for some examples when $\omega=0.8$: (A) $\nu=-4$  the phase space of initial conditions is broken in two, each one has a limiting $\theta$-periodic solution that are symmetric with respect the reflection $(q,\dot q)\mapsto -(q,\dot q)$ ; (B) $\nu=-2$ the stationary solution moves in a limited phase space region and (C) $\nu=-1.85$ where there is a numerically stable $2\theta$-periodic solution \cite{one2}.

 \begin{figure}[htbp] 
 \includegraphics[width=15cm]{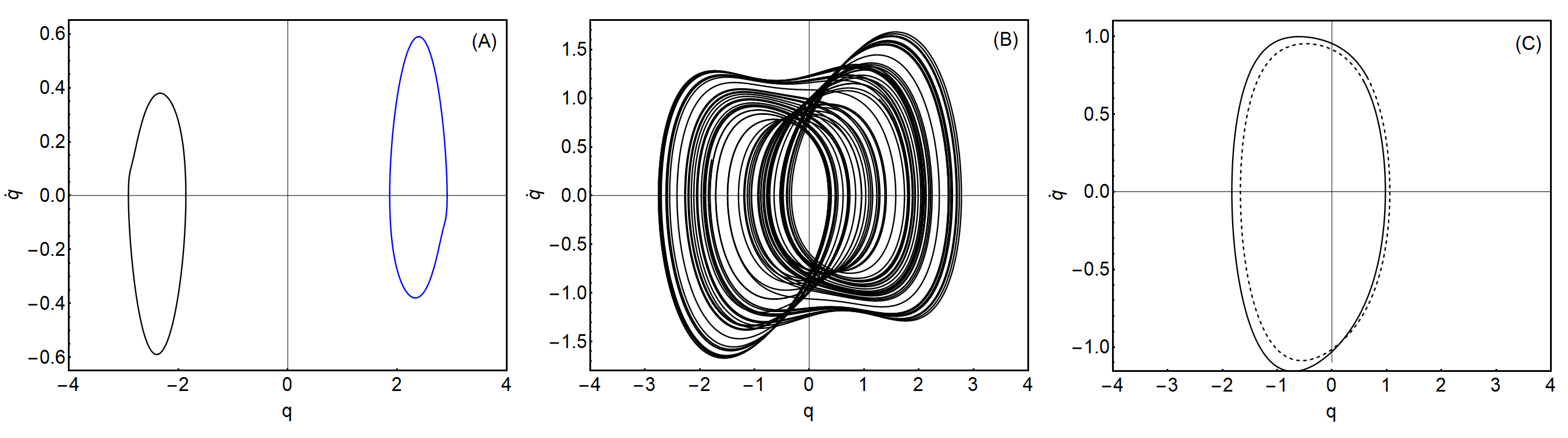} 
   \caption{Phase space plots $(q(t),\dot q(t))$ for $t\in(7000,7500)$ when $\omega=0.8$. (A) $\nu=-4$, (B) $\nu=-2$ and (C) $\nu=-1.85$}  \label{conf}       
 \end{figure}

\section{The many oscillator case}

 We solve the set of equations (\ref{eq:HZ100}) using a modified
 velocity-Verlet algorithm \cite{Allen}. Starting from a given initial
 condition we allow the system to relax over $10^7$ evolution steps of
 size $\Delta t=10^{-3}$. Any $\theta$-periodic observable, such as
 the work (\ref{013105-23}), is computed by sampling and summing
 equidistant time values  over one period of the force. This averaging
 process is repeated $10^6$ times. We take
 $\cal{F}(t/\theta)=F\cos(\omega t)$ with $\omega=2\pi/\theta$. In the
 simulations we use  $\omega_0=1$ and $\gamma=1$ and the anharmonic potential: $V(q)=q^4/(4(1+q^4))$ and
 $U(q)=0$. Observe that in this case
$\frak{V}=\frac{5}{16}\sqrt{\frac{5+\sqrt{5}}{2}}$. We have studied
$\omega=3$ with the radius of convergence of the perturbative expansion
$\bar\nu_0=6.7292\ldots$ and stability proven for
$\nu_0=0.1640\ldots$; $\omega=0.9$, with the radius of convergence $\bar\nu_0=0.3196\ldots$ but $\nu_0=0$ and, $\omega=0.5$ and $1.5$, where neither the convergence or the stability is known because some of its odd harmonics 
 are inside $\mathcal{I}$ in both cases.

\subsection{Relaxation to the stationary state:} It is rigorously
   known that for the harmonic case, $\nu=0$,  and fixed $N$, $F$ and
   $\gamma>0$, starting from any bounded initial condition the system
   approaches, as $t\rightarrow\infty$,   a unique
     $\theta$-periodic state. Moreover, it has been shown that if
   $F=0$, $\gamma>0$, then the system of size $2N+1$ relaxes to the stationary state
   at the rate $\exp(-A t/N^3)$, when $t\rightarrow\infty$
   \cite{menegaki}. The constant $A$ does not depend on $N$.

 We have made several computer simulations to verify this behavior
for $F=0$. In particular we show in Figure \ref{relax} how the energy
per particle, $u(t)= \frac{1}{N}\mathcal{H}_N (\mathbf q(t), \dot{\mathbf q}(t)\big)$, goes to zero as
$t\rightarrow\infty$, which is consistent with
\eqref{060810-24}. 
The initial condition is $\dot q_x(0)=0$, $q_x(0)=\cos(8\pi x/(2N+1))$, $x\in\mathbb{Z}_N$, $\omega_0=1$. We study the cases $\nu=0$ and $\nu=1$ for $N=50$, $100$, $150$ and $200$. 
 \begin{figure}[htbp] 
 \includegraphics[width=6.5cm]{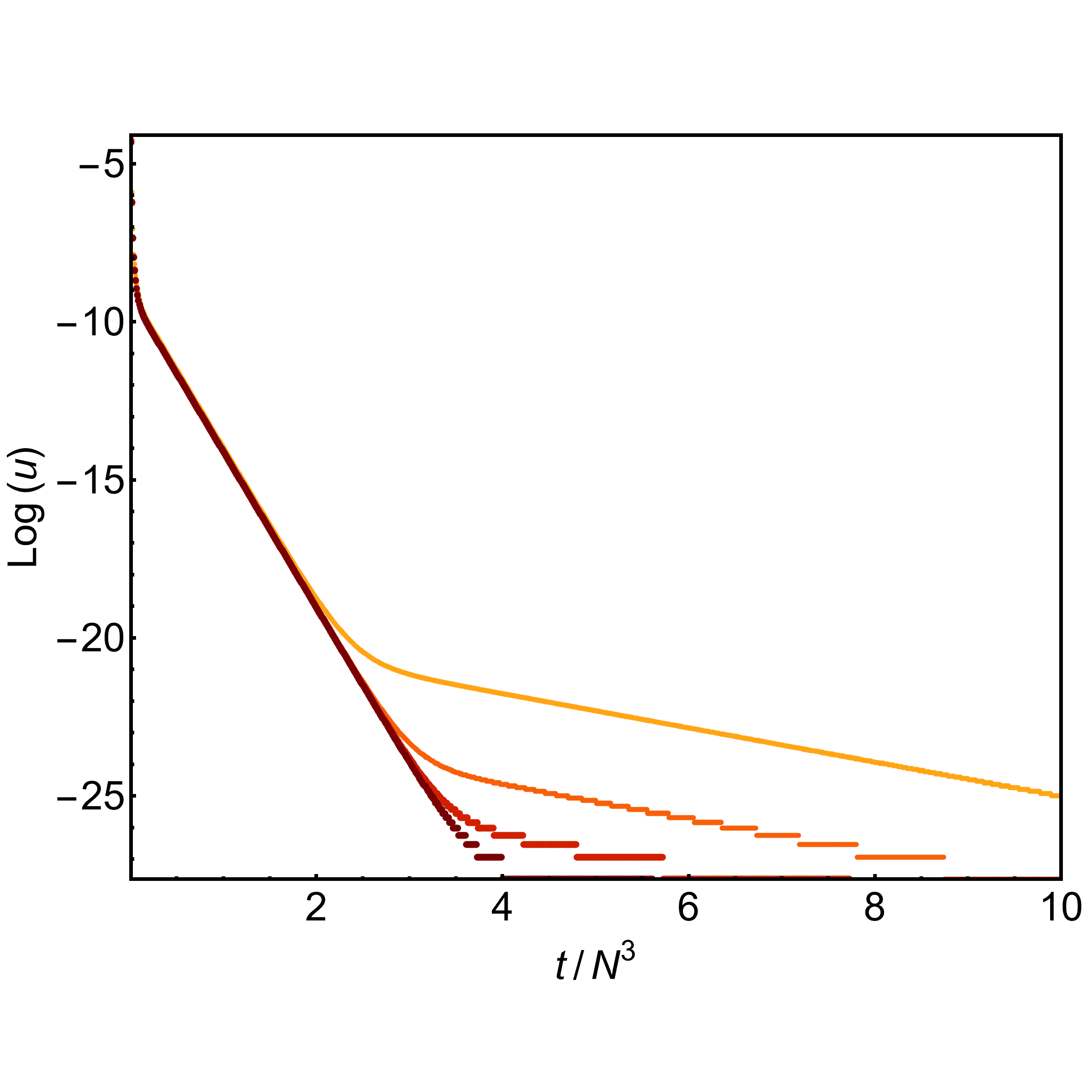} 
 \includegraphics[width=6.5cm]{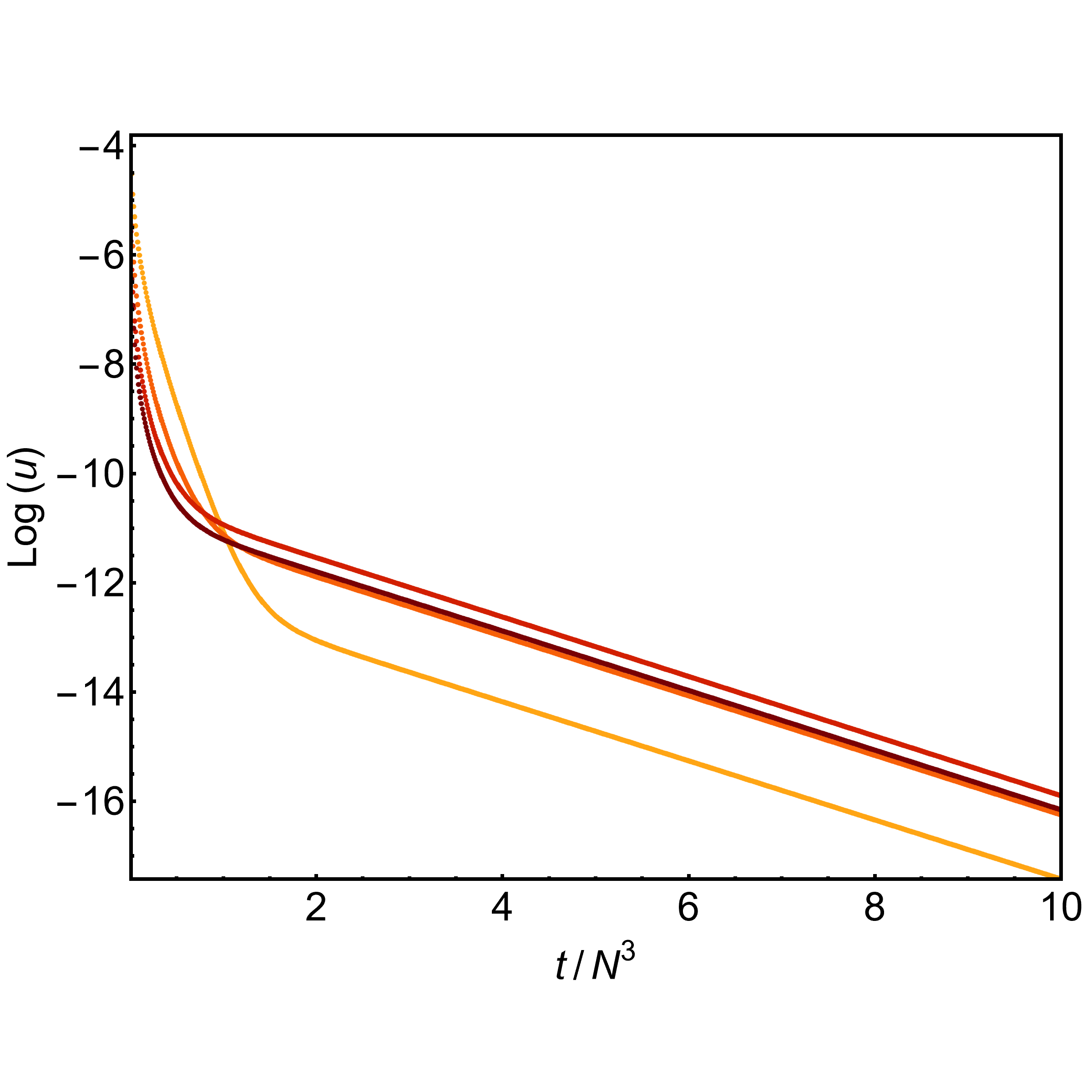} 
 \caption{$u$ vs. $t/N^3$ ($F=0$, $\omega_0=1$) for $\nu=0$ (left figure) and $\nu=1$ (right figure). $N=50$ (yellow points), $N=100$ (orange points), $N=150$ (red points) and $N=200$ (dark-red points).} \label{relax}        
\end{figure}

 We observe that for $\nu=0$ (the harmonic case) there is a perfect scaling of all $N$-s for $\tau=t/N^3>0.1$ up to  a value that increases with $N$, where finite size effects appear. For $\nu=1$, the long time decay looks again as $B(N)\exp(-A t/N^3)$ with $
 A$ independent on $N$ and it seems that $B(N)\rightarrow B$. Finally
 $A(\nu=0)>A(\nu=1)$, that is, the anharmonicity slows the decay to
 the energy  equilibrium.

 \subsection{Spatial behavior of the $\theta$-periodic stationary state
     for $F\neq 0$:} From the numerical analysis it seems that, as in
   the harmonic case, the large scale spatial behavior of the
     Fourier modes  $\tilde
   q_x(m)=\theta^{-1}\int_0^\theta e^{-imt}q_{x,{\rm p}}(t)dt$}  of the periodic solution $q_{x,{\rm
       p}}(t)$, depends on whether $m\omega$ -  some integer
 multiplicity of $\om$ - lies within the harmonic interval or not. In
particular, we observe that for fixed $\nu$, $\gamma$ and $N\gg 1$:
 \begin{itemize}
  \item $m\omega\notin\mathcal{I}\Rightarrow \tilde q_x(m)\simeq e^{-c(m\omega)\vert x\vert}$ 
  \item  $m\omega\in\mathcal{I} \Rightarrow \tilde q_x(m)\simeq e^{i k(m\omega)\vert x\vert}$
 \end{itemize}
where $c$ and $k$ are real positive valued functions and $\mathcal{I}=[1,\sqrt{5}]$.

\begin{figure}[htbp] 
 \includegraphics[width=6.5cm]{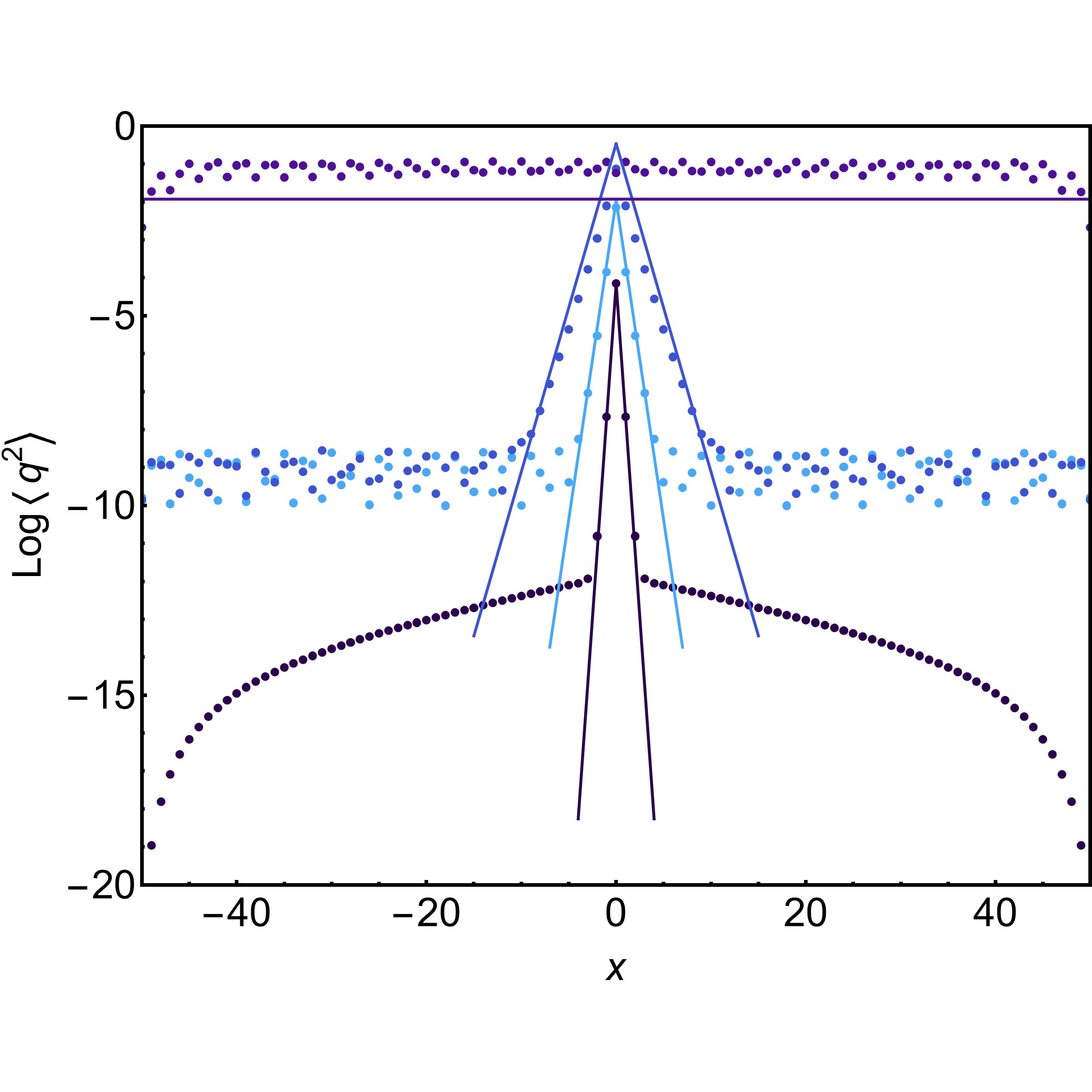} 
 \includegraphics[width=6.5cm]{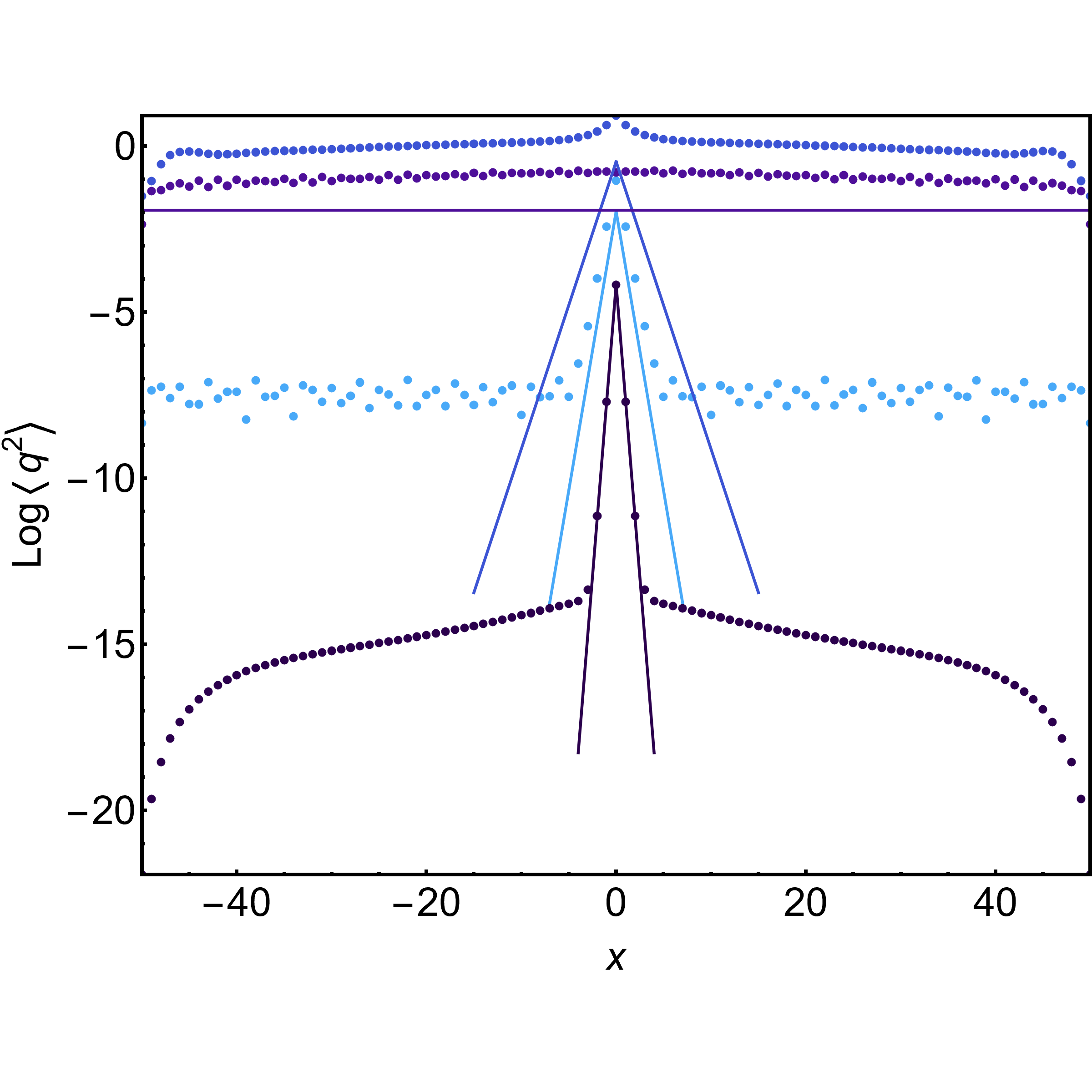} 
 \captionof{figure} {$\langle q_{x,{\rm p}}^2\rangle$ vs. $x$ for $F=1$ and $\omega=0.5$ (cyan points), $\omega=0.9$ (blue points), $\omega=1.5$ (violet points) and $\omega=3$ (black points). Left: $\nu=1$ and Right: $\nu=-2$. Lines show the exact solution for each $\omega$ for the infinite harmonic case ($\nu=0$, $N\rightarrow\infty$).
 } \label{fin7}        
 \end{figure}

To illustrate this behavior, we plot in Figure \ref{fin7}   the quantity
$\langle q_{x,{\rm p}}^2(t) \rangle$ - the
square of the amplitude  $q_{x,{\rm p}}^2(t)$   averaged  over one period $\theta$. 
In Figure \ref{fin7} $F = 1$, $N=50$ and $\omega = 0.5$, $0.9$,
$1.5$, and $3$, with $\nu = 1$ (left figure) and $\nu = -2$ (right
figure).  Observe that only the case $\nu=1$ and $\omega=3$ could be studied by our perturbative scheme.

Generally, we observe an exponential decay around the origin whenever $\omega \notin \mathcal{I}$. 
It is noteworthy that for $\omega = 3 > \sqrt{5}$, the overall behavior appears similar to the harmonic case ($\nu = 0$) for both values of $\nu$ studied.
When $\omega<1$ we see a non-exponential decay far from the origin due to finite size effects that tend to dissapear as $N$ increases. Furthermore, when $\omega \in \mathcal{I}$, we observe an average constant value along $x$ with some boundary effects near the lattice end points.  This is consistent with the possibility that, for $\nu \neq 0$, there are plane waves traveling from the origin to $\pm \infty$, as in the infinite harmonic case. Finally, we highlight in Figure \ref{fin7} the singular behavior for $\omega = 0.9$ and $\nu = -2$, where the amplitudes exhibit a smooth decay as we move away from the origin, which is compatible with a power-law decay. This behavior is related to the known phenomenon of supratransmission \cite{Geniet,supra}.

 In order to minimize finite size effects and understand the system's
 behavior for $N>>1$ we have performed computer simulations for a
  large system  with $N=10^4$,
 $F=1$. Starting with the initial condition: $q_x(0)=\dot q_x(0)=0$ we
 let the system evolve. When any of the particles at $\pm N/2$ moves
 for the first time, we begin to record $q_x(t_n)$ at times
 $t_n=n\Delta t$, $n=0,1,...$, for $x=0,1,\ldots,40$, $\Delta
 t=2\pi/(100\omega)$. We stop recording when the particles at the end,
 $q_{\pm N}$, begin to move to discard boundary effects. We analyze the spatial and dynamic structure of the data files that have an extension of $N_D=6\times 10^4-4\times 10^5$ depending on $\omega$. We check that the observables are stationary by analyzing different time intervals. 

\begin{figure}[htbp] 
 \includegraphics[width=7cm]{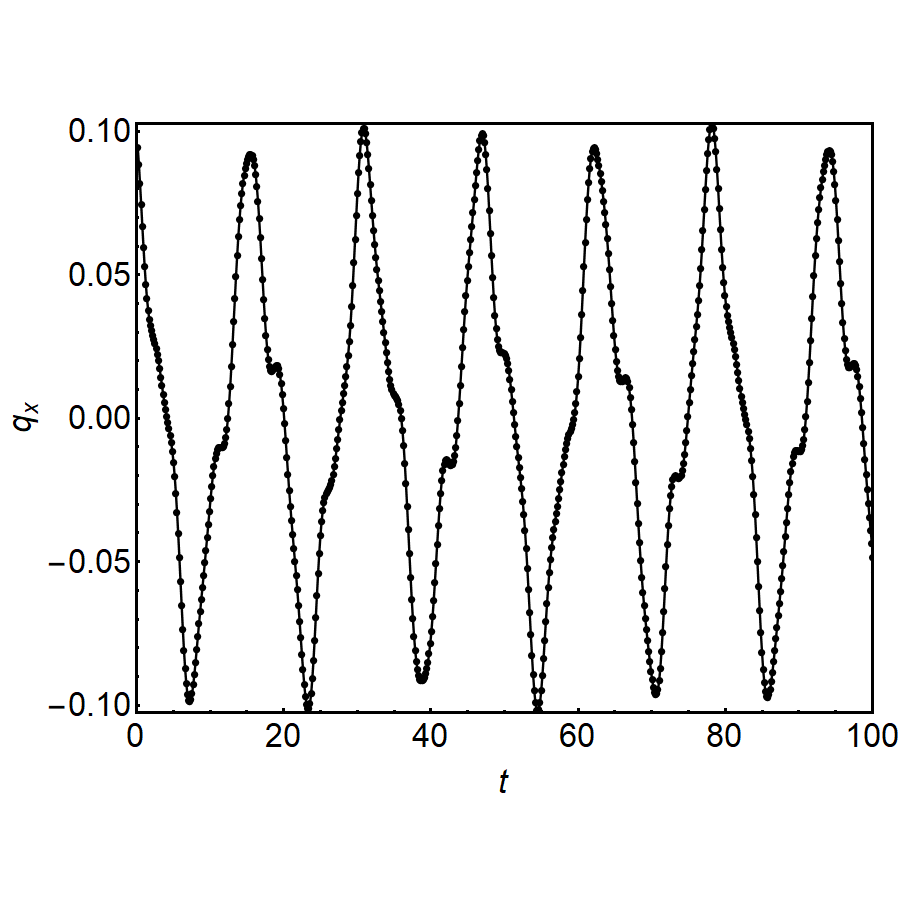} 
 \caption{$q_x(t)$ vs. $t$ obtained from the dynamic simulation (see text) with $N=10^4$, $\nu=1$, $\omega=0.4$ and $x=3$. 
 } \label{fin10}        
 \end{figure}

\begin{figure}[htbp] 
\vskip -0.5cm
 \includegraphics[width=7cm]{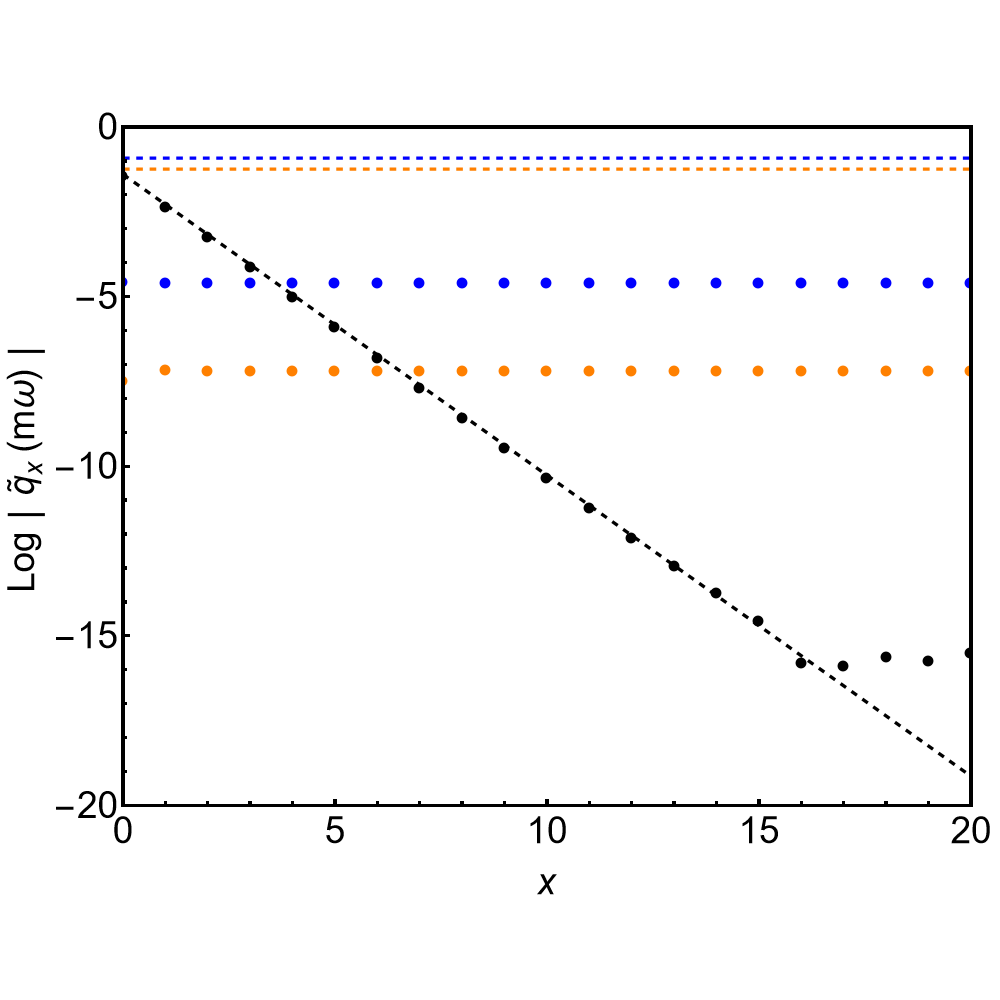} 
 \caption{$\log\vert \tilde q_x(m\omega)\vert$ vs. $x$ obtained from the dynamic simulation (see text) with $N=10^4$, $\nu=1$, $\omega=0.4$. $m=1$ (Black dots), $m=3$ (Blue dots) and $m=5$ (Red dots). Dashed lines correspond to the exact solution for the harmonic case when $m=1$ (Black line), $m=3$ (Blue line) and $m=5$ (Red line).   
 } \label{fin11}        
 \end{figure}

\begin{figure} 
\hskip -2cm
  \includegraphics[width=15cm]{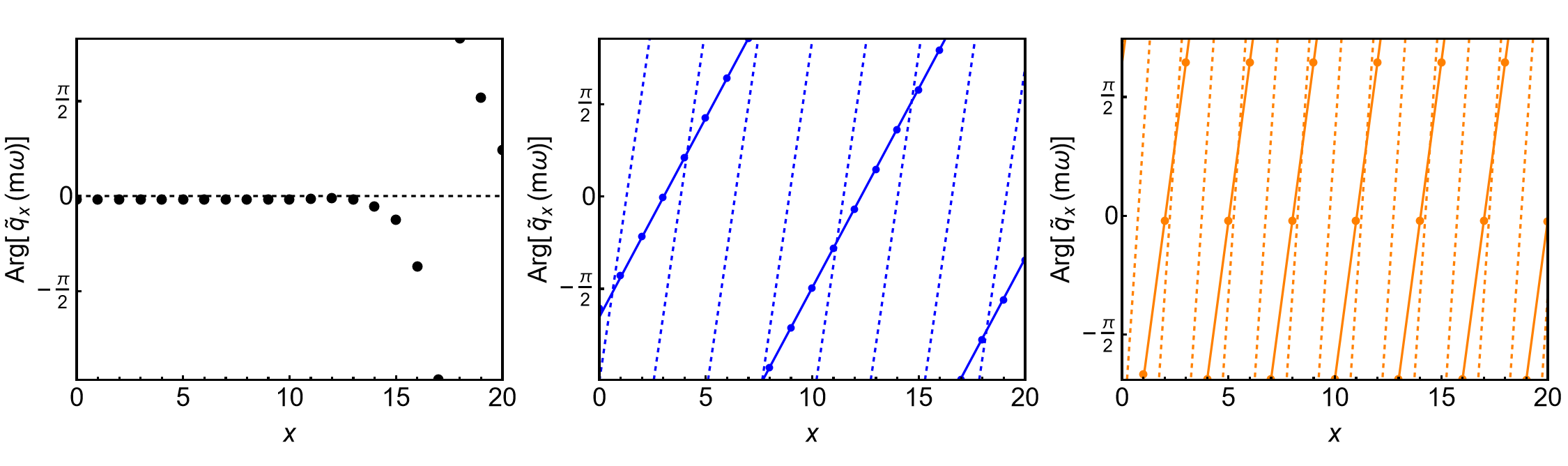} 
 \caption{$\text{Arg}(\tilde q_x(m\omega))$ vs $x$  obtained from the dynamic simulation (see text) with $N=10^4$, $\nu=1$, $\omega=0.4$. $m=1$ (Black dots), $m=3$ (Blue dots) and $m=5$ (Red dots).Dashed line corresponds to the values obtained for the harmonic case.   
 } \label{fin12}        
 \end{figure}

We use the values of $\omega$ such that some of its odd multiplicities lie in $\mathcal{I}$ and, therefore, our perturbation theory cannot be applied to them. 
  In Figure \ref{fin10} we show an example of a typical evolution:
  $\omega=0.4$ and $x=3$. We see the non-trivial periodic
  evolution. The dynamic sequence recorded \sout{it} is Fourier
  transformed: $\tilde q_x(-\pi/\Delta t+2\pi s/(N_D \Delta t))$,
  $s=1,\ldots,N_D$. We show in Figures \ref{fin11} and \ref{fin12} how
  $\tilde q_x(m\omega)$ decay with $x$ for $m=1$, $3$ and $5$. We see
  that the first harmonic $m=1$ decays exponentially with $x$ and
  it matches the behavior of the infinite harmonic case
  (the dotted line). However, the harmonics $m=3$ and $m=5$
  that are inside  interval $\mathcal{I}$
correspond  to constant amplitude and
periodic constant of the argument of $\tilde
   q_x(m)$. This indicates the existence of a plain wave structure
 but far from the values corresponding to the harmonic solution. In
 fact, its corresponding slopes, $k$, are smaller than their harmonic counterparts $k_h$: $k(m=3)=0.678$, $k_h(m=3)=2.5$ and $k(m=5)=2.095$, $k_h(m=5)=4.285$.


 \subsection{The work.}  We have measured $W$, see
     \eqref{013105-23}, for different values of $N$, $F$ and
     $\gamma>0$. The simulations indicate that, as
     $N\rightarrow\infty$, the work functional   generically behaves as follows:
 \begin{itemize}
 \item  ${\omega<\min\mathcal{I}}$: $ {W\simeq 0}$ when
   ${\nu\geq 0}$ and ${W>0}$  for  $\nu$ negative enough,
 \item ${\omega\in \mathcal{I}}$: ${W\neq 0}$, regardless of the
     sign of $\nu$,
 \item  ${\omega>\max\mathcal{I}}$: ${W\simeq 0}$, regardless of the
     sign of $\nu$. 
 \end{itemize}

\begin{figure}[htbp]
 \includegraphics[width=6.5cm]{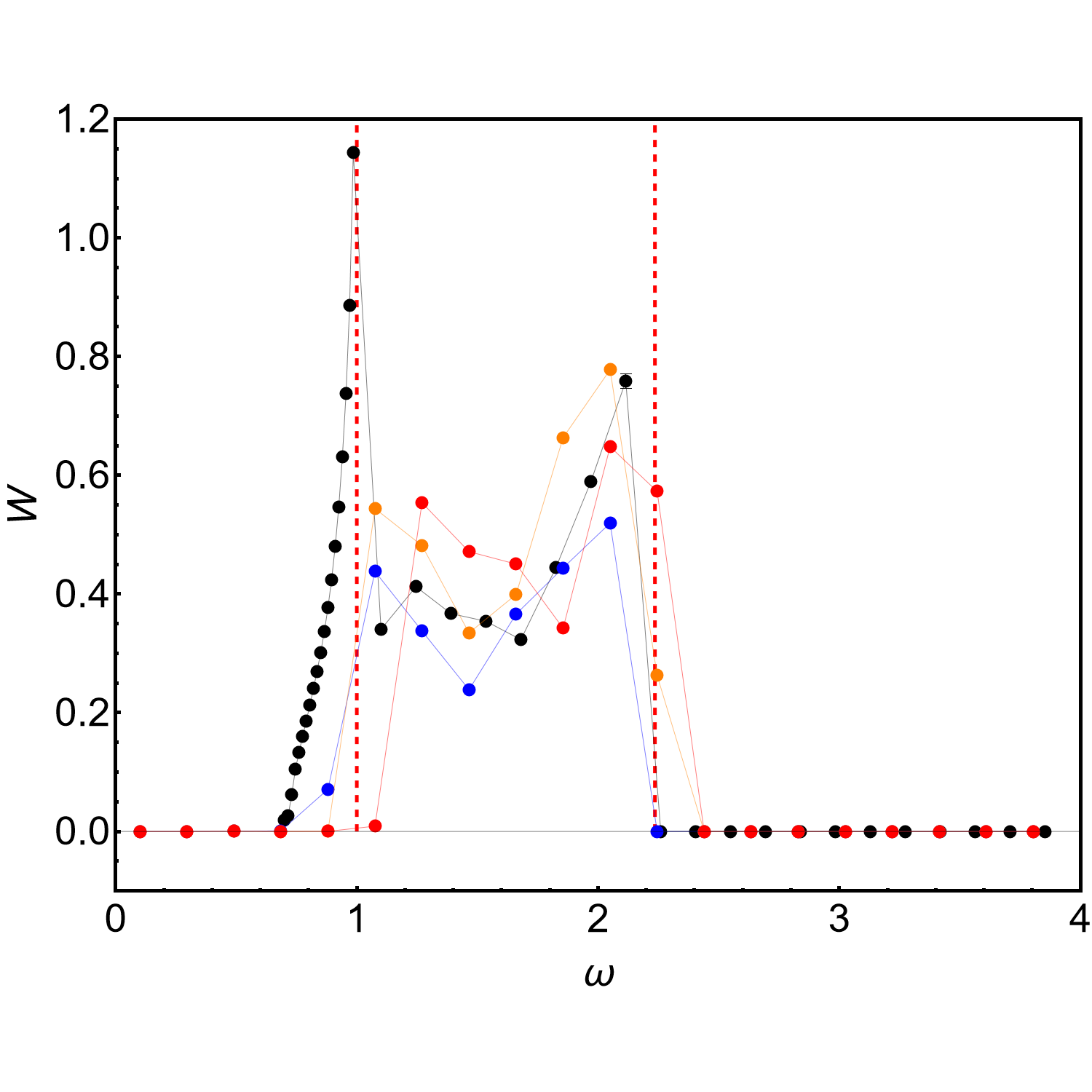} 
 \includegraphics[width=6.5cm]{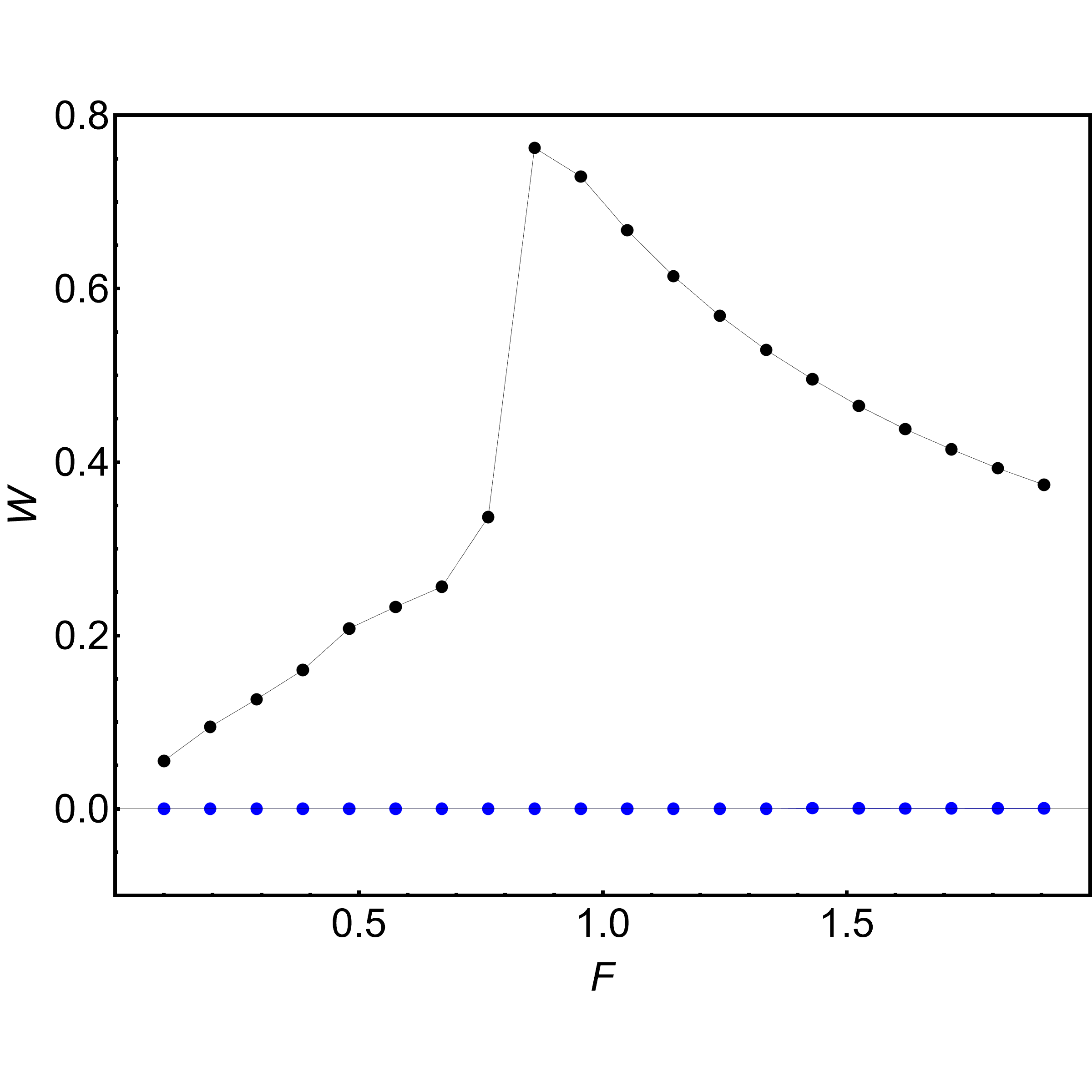} 
 \caption{Left: Work vs. $\omega$ for $F=1$ and $N=50$. $\nu=-2$ (black points), $\nu=-1$ (blue points), $\nu=1$ (orange points), $\nu=2$ (red points). Red dashed lines show the limit of the harmonic interval. Right: Work vs. $F$ for  $N=50$,  $\nu=-2$ and $\omega=0.95$ (black points) and $\omega=3$ (blue points). } \label{fin3}        
 \end{figure}

 Figure \ref{fin3} (right) shows an example of the behavior of the
 work. There $N = 50$, $\omega_0 = 1$, and $\gamma = 1$ and we plot
 $W$ as a function of $\omega$ for $\nu = -2$, $-1$, $1$, and $2$. We
 observe that the work is almost zero for all values of $\nu$ when
 $\omega > \mathcal{I}$. Inside the interval $\mathcal{I}$, the work
 is nonzero and fluctuating (as we previously observed in the harmonic
 case, $\nu = 0$). For $\omega <\min\mathcal{I}$ and $\nu>0$ the work
 is zero but when $\nu<0$ and sufficiently large,  a new
 phenomenon emerges. Specifically, for $\nu=-2$ and $F=1$ the
 forcing frequencies $\omega$  in the interval $[0.68,1]$ result in
 non-zero work, even though they lie outside the harmonic interval,
 $\mathcal{I}$. This is  known as the supra-transmission
   phenomena, see \cite{Geniet,supra}. We also observe in Figure
 \ref{fin3} that in the supra-transmission regime the force does not play an important role. For instance, when $\omega=0.95$ and $\nu=-2$,  the work reaches its maximum for $F\simeq 1$ and it decays to zero for large $F$-s.

\begin{figure}[htbp]
 \includegraphics[width=6.5cm]{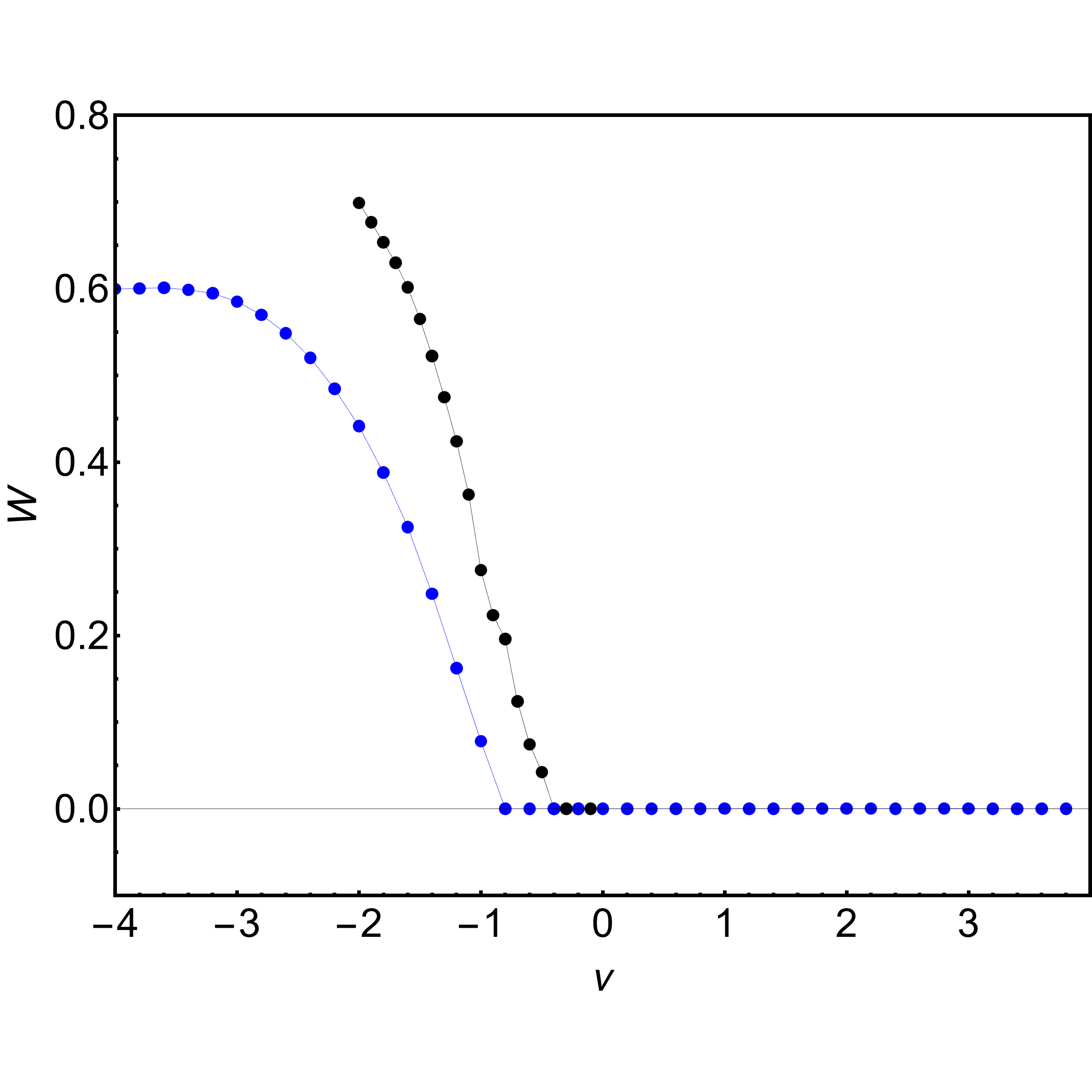} 
 \includegraphics[width=6.5cm]{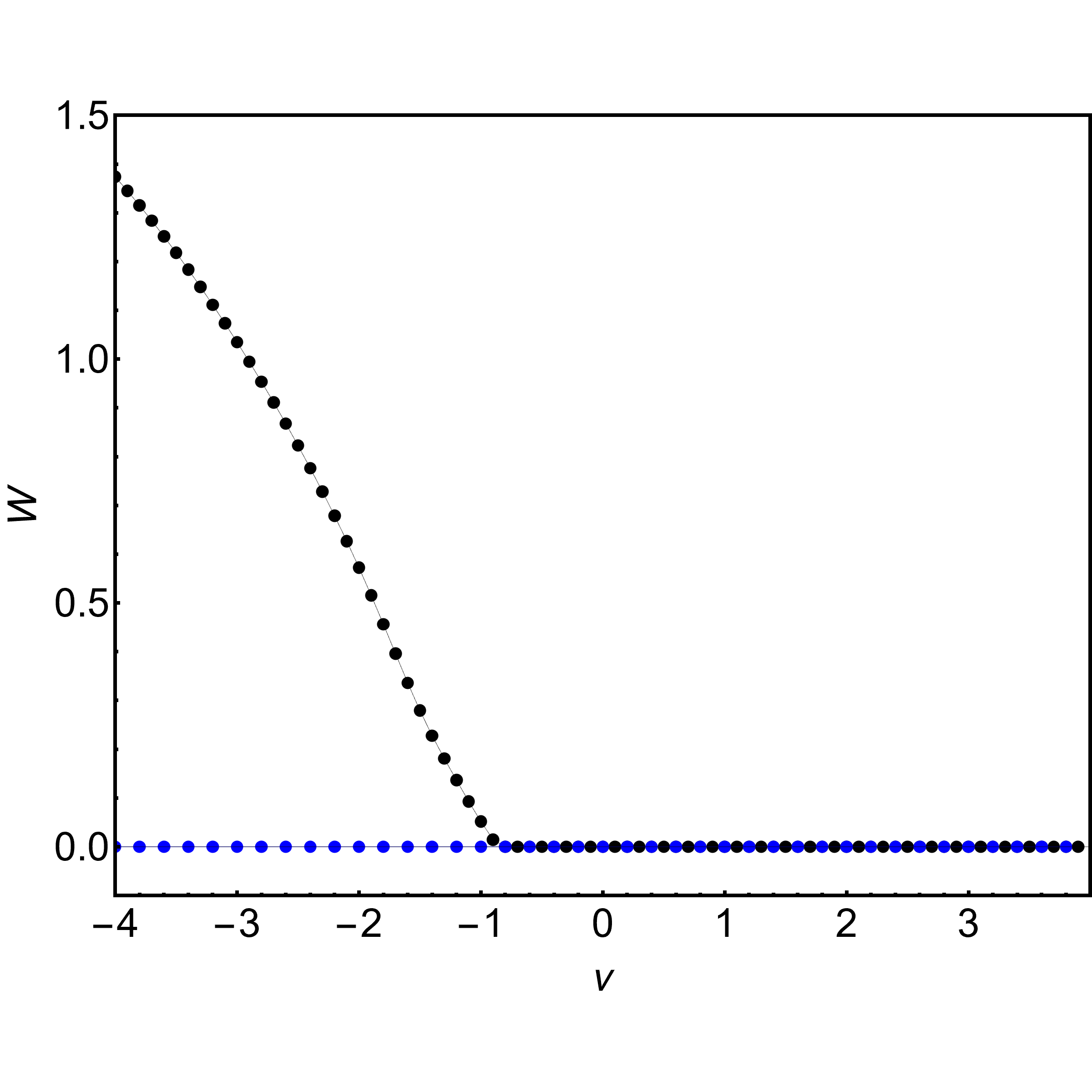} 
 \caption{Work vs. $\nu$ for  $F=1$ and $N=50$. Left: $\omega_0=1$, $\omega=0.95$ (black points) and $\omega=0.9$ (blue points). Right: $\omega_0=2$, $\omega=1.95$ (black points) and $\omega=1$ (blue points). 
 } \label{fin5}        
 \end{figure}

 In Figure \ref{fin5} we plot work versus $\nu$ for some $\omega$ values:  $\omega=0.9$  and $0.95$ with $\omega_0=1$ (left figure) and $\omega=1$ and $1.95$ with $\omega_0=2$ (right figure). We see that the supratransmission regime , $W>0$, appears in both cases for $\nu<\nu_c(\omega)<0$ where $\nu_c(0.95)\simeq -0.3$, $\nu_c(0.9)\simeq -0.8$ when $\omega_0=1$ and $\nu_c(1.95)\simeq -0.87$ when $\omega_0=2$.

\section{Bounded vs unbounded potentials}

The perturbation theory described in the present article relies
fundamentally on the assumption that both pinning and interacting
potentials have bounded second derivatives. We stress that potentials with unbounded second derivatives are not covered by our theory. In this section, we compare the behavior at the stationary state for both bounded and unbounded potential cases. 
We are particularly interested in the cases studied by Prem et
al. \cite{Prem} and by Kumar et al.\cite{Kumar} that correspond to
$U(q)=0$ and $V(q)=q^4/4$. They found that the work done on the system
by the external force $F$ has a surprising behavior for large $F$ when
$\omega$ is inside the harmonic spectrum. In particular for
fixed $N$ and $\gamma$ the work becomes independent of
$F$ for some interval and then it decreases with $F$. To see if this phenomenon depends on the potential we performed simulations for
 $N=50$, $\omega_0=1$, $\gamma=1$, $\omega=0.8$ and $1.5$ (the former
 forcing frequency lying   outside and the latter inside the
   harmonic interval) as well as for $F\in[0,6]$ and $\nu=\pm 0.3$,
 $\pm 0.6$ and $1$. Observe that for $\omega=0.8$ we have
 $\bar\nu_0=0.60563$ and the stability bound of our theory, in the
 bounded potential case as in \eqref{012506-25}, does not apply
because some multiplicity of $\omega$ lies in the harmonic interval. The potentials considered are:
 \begin{equation}
   \label{012506-25}
\text{bounded:}\quad V(q)=\frac{q^4}{4(1+q^4)}\quad,\quad\text{unbounded:}\quad V(q)=\frac{q^4}{4}
\end{equation}

\begin{figure}[htbp]
 \includegraphics[width=6.5cm]{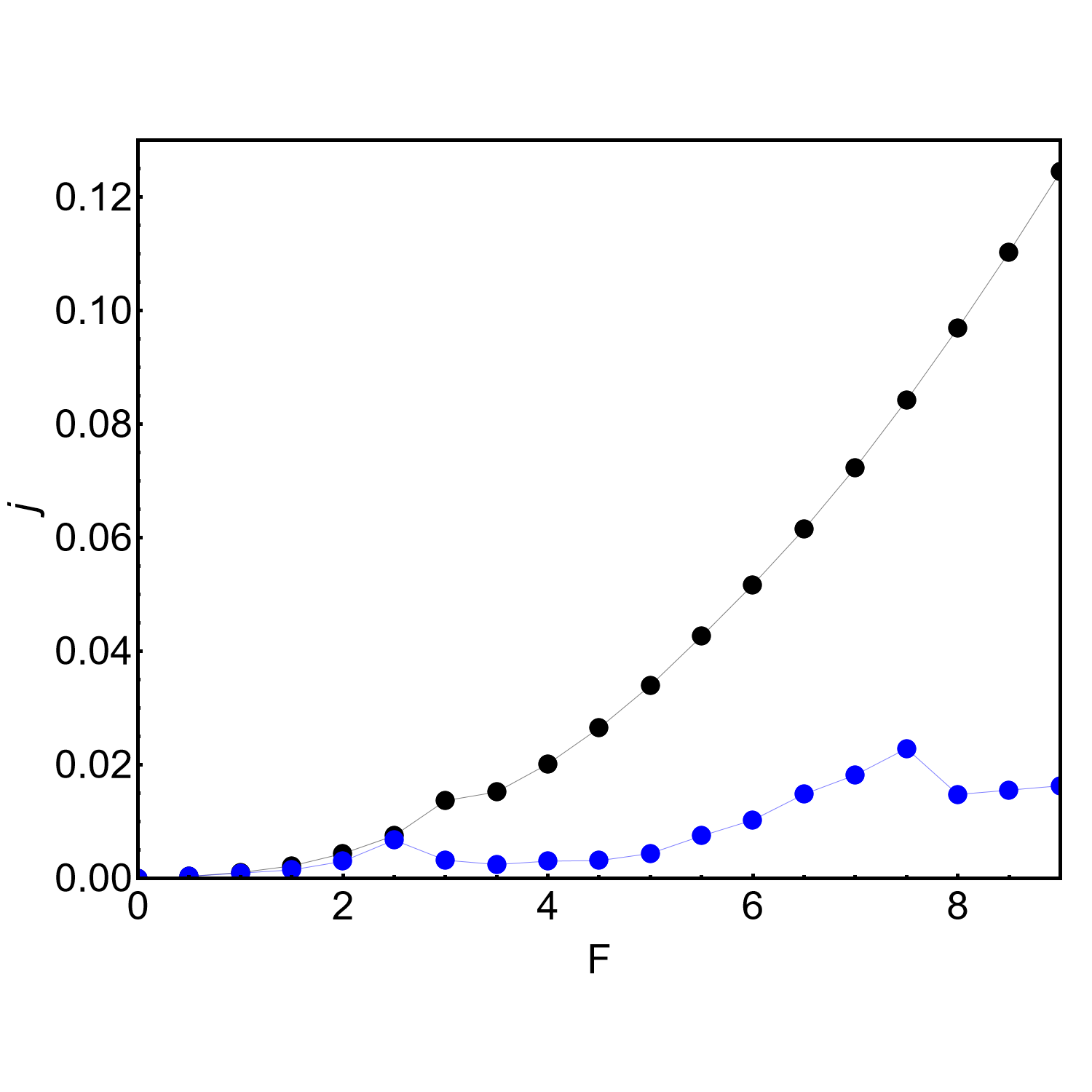} 
 \includegraphics[width=6.5cm]{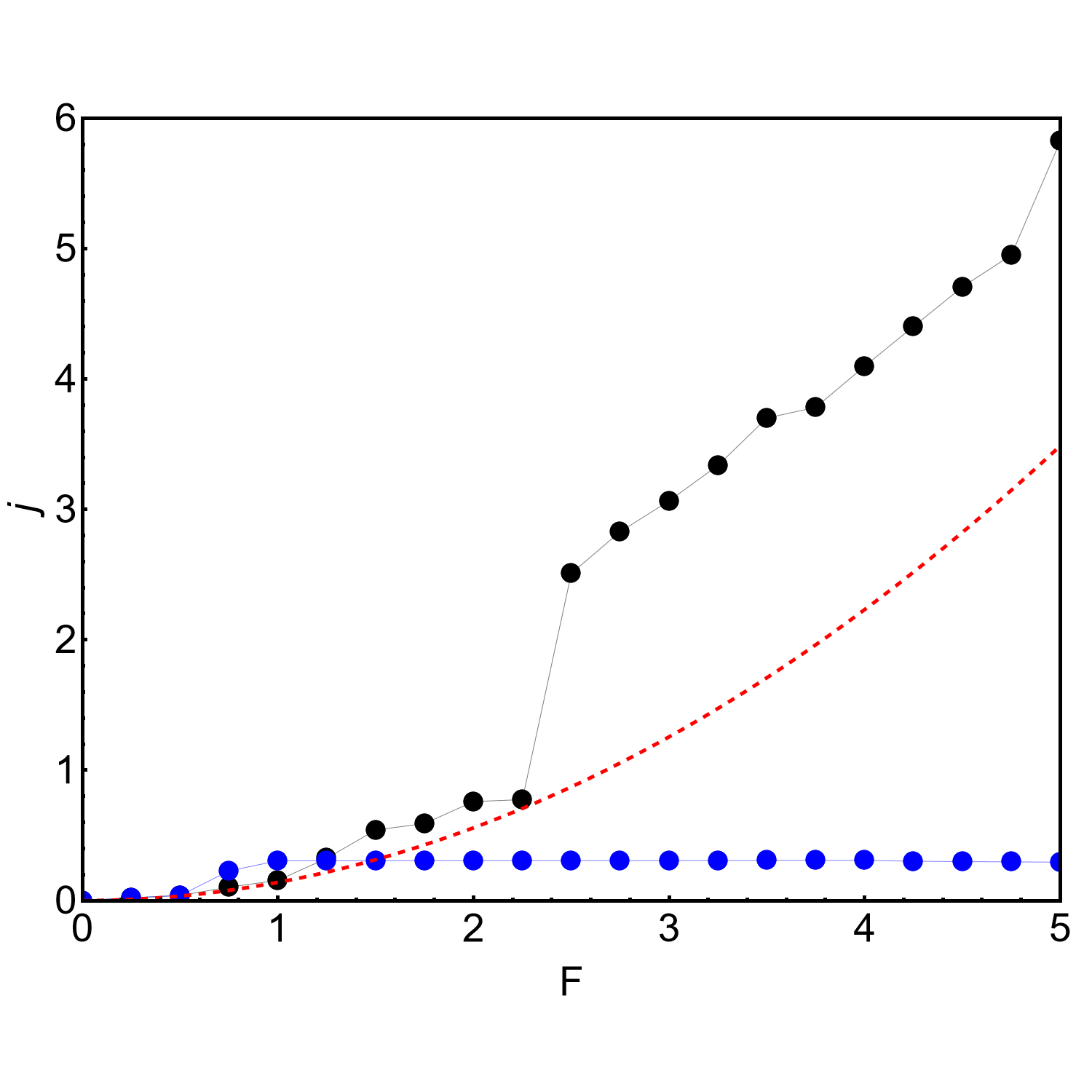} 
  \caption{Averaged energy current $j$ vs. $F$ for $\omega=0.8$ (left) and $\omega=1.5$ (right). Black points are for the bounded potential $V(q)=q^4/(4(1+q^4))$ and blue points for the unbounded potential $V(q)=q^4/4$. Red dashed line is the current for the harmonic case for $N=50$. Observe that for $\omega=0.8$ the current for the finite harmonic case is of order $\simeq 10^{-25}$ for the interval of $F$'s in the figure.
 } \label{co}        
 \end{figure}

Figure \ref{co} shows the energy current $j$, averaged over space and one period of the external force,
 for $\omega=0.8$ and $\omega=1.5$, plotted against the force amplitude
 $F$.  We have $j=W/2$, as the current carries the work done by $F$ to both ends of the chain. For the bounded potential, we observe that for large values of $F$, $j\simeq F^2$ when $\omega=0.8$ and $j\simeq F$ when $\omega=1.5$.
In contrast, for the unbounded potential, the averaged current appears to saturate, approaching a constant value, as 
$F$ increases in both frequency cases.

Additionally, we observe distinct values of the forcing amplitude $F$ at which the current behavior changes. For the bounded case, these transitions occur at approximately $F\simeq 3$
  for $\omega=0.8$ and, $F\simeq 2.5$ for $\omega=1.5$. In the unbounded case, they occur at $F\simeq 2.5$  for $\omega=0.8$ and, $F\simeq 1$  for $\omega=1.5$. We believe these features are due to finite-size effects.

\begin{figure}[htbp]
 \includegraphics[width=14cm]{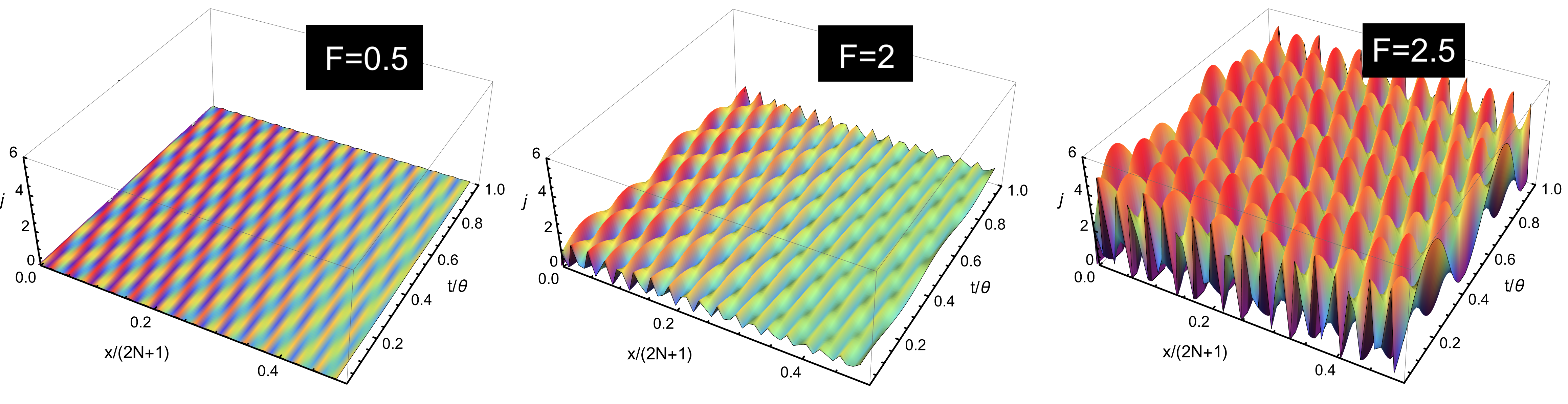} 
 \includegraphics[width=14cm]{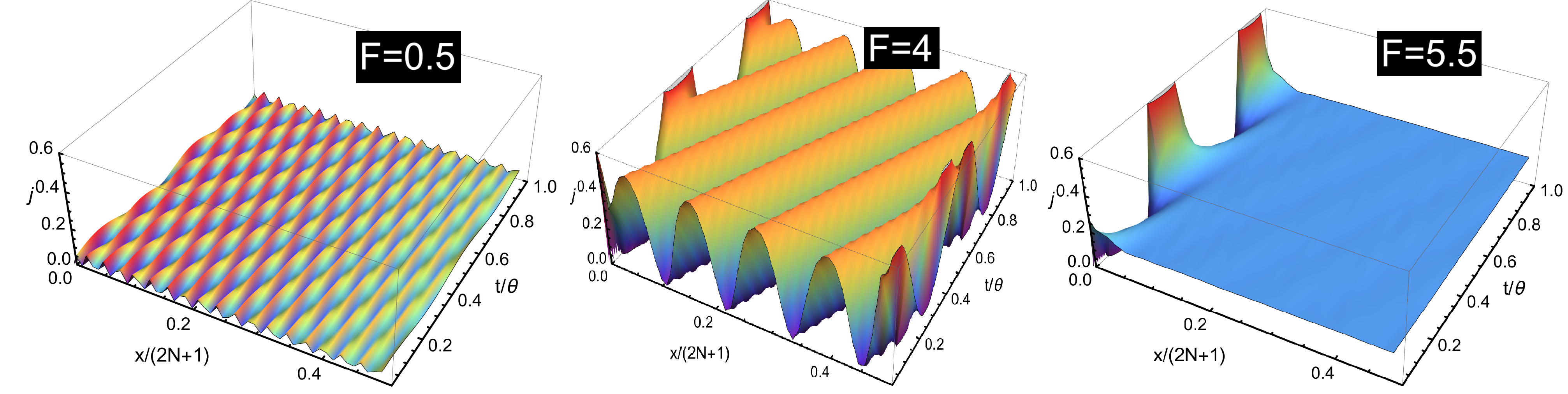} 
  \caption{$j$ vs. $(x/N,t/\theta)$ for $\omega=1.5$. First row are for the bounded potential $V(q)=q^4/(4(1+q^4))$ and $F=0.5, 2$ and $2.5$ (from left to right) and the second row  are for the unbounded potential  $V(q)=q^4/4$ and $F=0.5$, $F=4$ and $F=5.5$ (from left to right).
 } \label{com2}        
 \end{figure}

Figure \ref{com2} presents a more detailed view of the spatio-temporal
behavior of the current. For low values of $F$, both potentials
exhibit similar current profiles in both space and time. In the case of the
bounded potential, the amplitude of the waves, as well as their
average over both space and time, increases monotonically with
$F$, while the overall spatio-temporal structure remains the
  same (see the first row, corresponding to $F=0.5$ of Figure \ref{com2}).  

In contrast, the unbounded potential shows qualitatively different behaviors depending on the value of 
$F$. For example, we observe plane wave structures that remain unchanged in the interval $F\in[1,5]$, followed by a transition to a spatially and temporally constant current for $F>5$. We believe this behavior may be attributed to strong interactions with the system boundaries in the unbounded potential case.

\begin{figure}[htbp]
 \includegraphics[width=14cm]{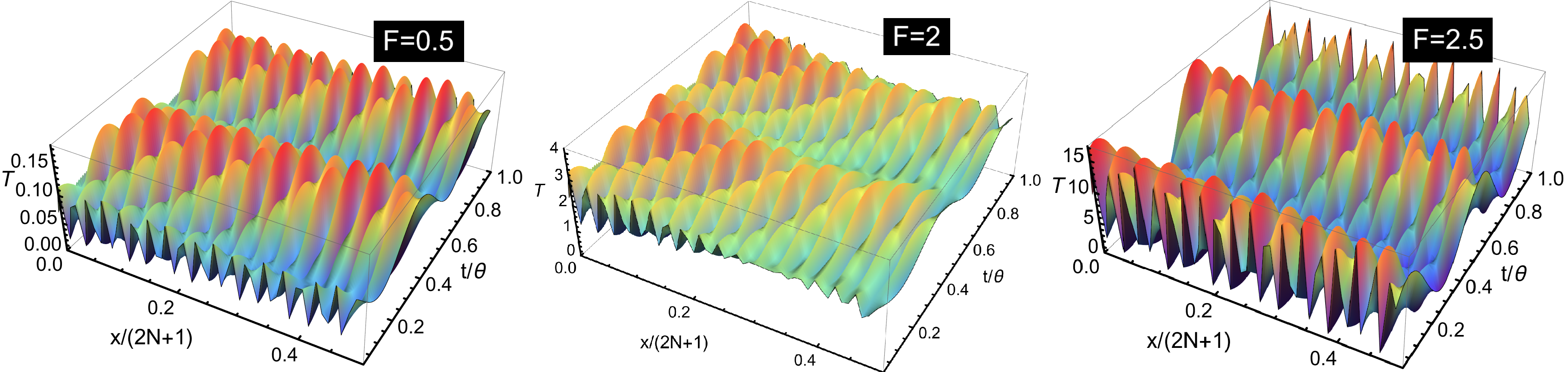} 
 \includegraphics[width=14cm]{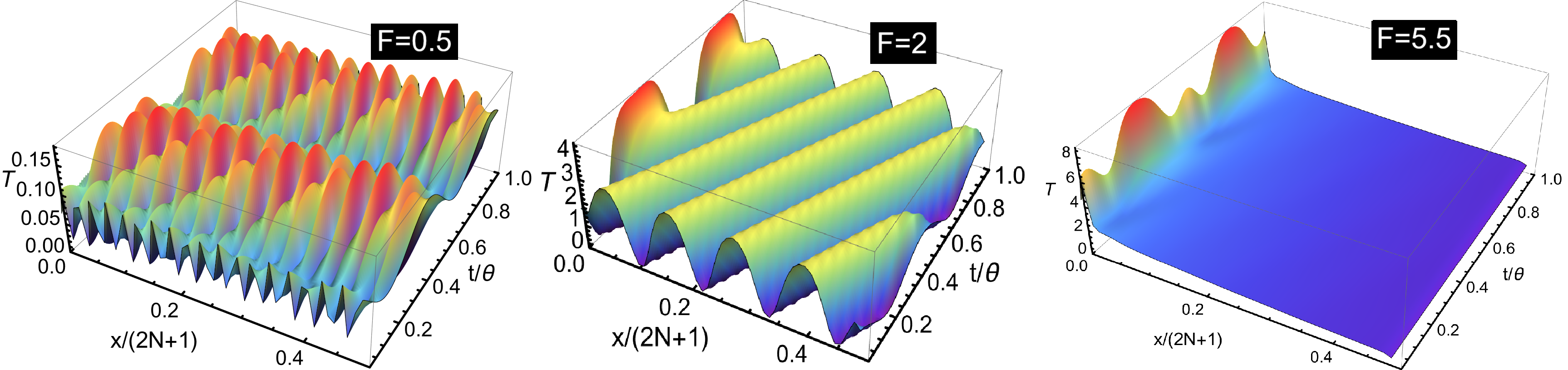} 
  \caption{$T$ vs. $(x/N,t/\theta)$ for
    $\omega=1.5$. First row are for the bounded potential
    $V(q)=q^4/(4(1+q^4))$ and $F=0.5, 2$ and $2.5$ (from the left to right) and the second row  are for the unbounded potential  $V(q)=q^4/4$ and $F=0.5$, $F=2$ and $F=5.5$ (from the left to right).
 } \label{tem}        
 \end{figure}

Finally, we show in Figure \ref{tem} the temperature $T$ interpreted as  $\langle \dot q_x^2\rangle$, of each oscillator at the stationary state. We see again no difference between the bounded and unbounded cases for small values of $F$. For the bounded case, the pattern changes from $F=2$ to $F=2.5$. Observe how the temperature concentrates  in most of the oscillators at the same time: $t\simeq 0.4\theta$ for $F=0.5$ or $t\simeq 0.6\theta$ for $F=2.5$ for the bounded potential case. For the unbounded case it is \sout{again} remarkable to see again the plain waves when $F=2$ and a constant temperature profile (independent on $t$ for $F=5.5$ that is consistent with the constant current observed in such case.

\begin{figure}[htb]
 \includegraphics[width=7cm]{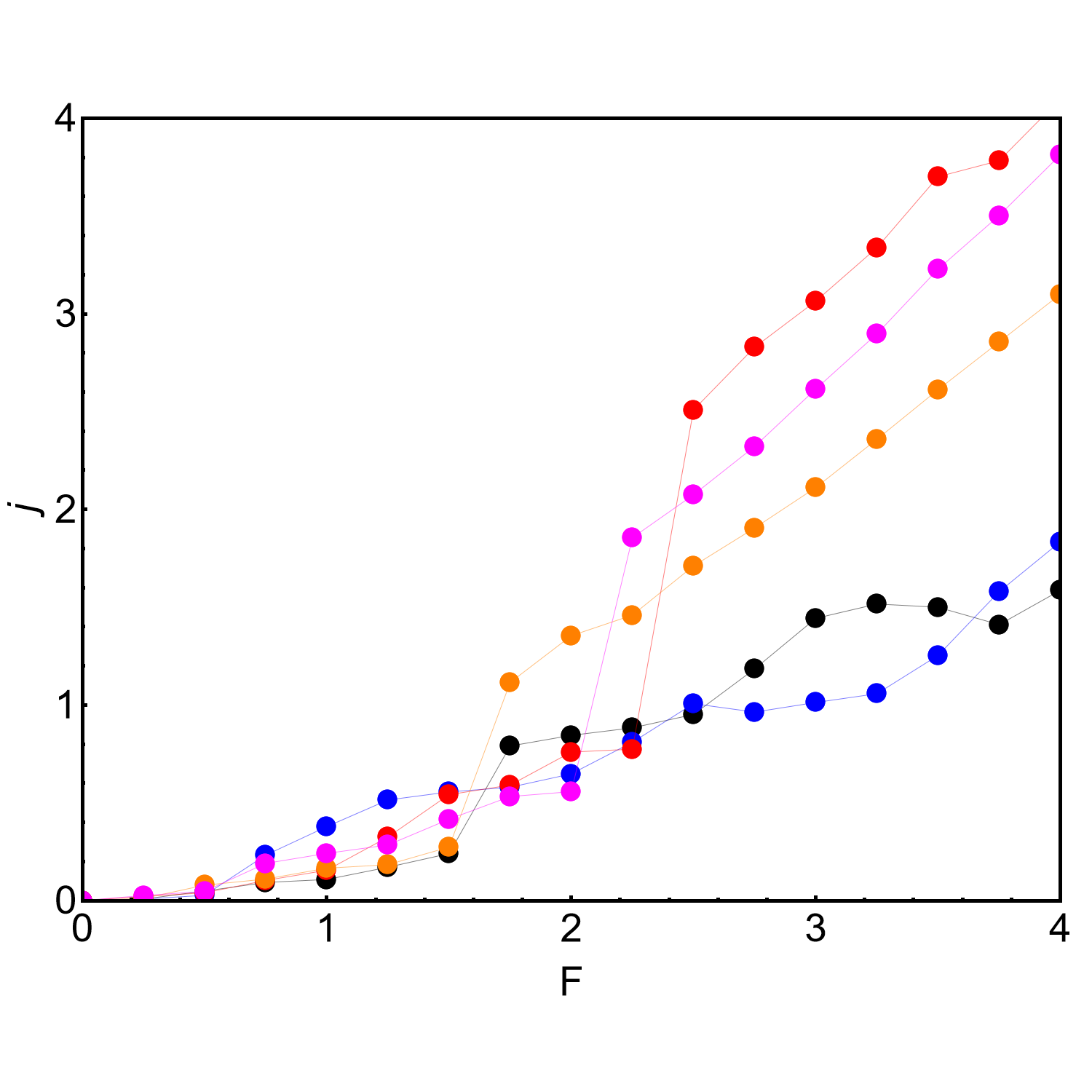} 
 \includegraphics[width=7cm]{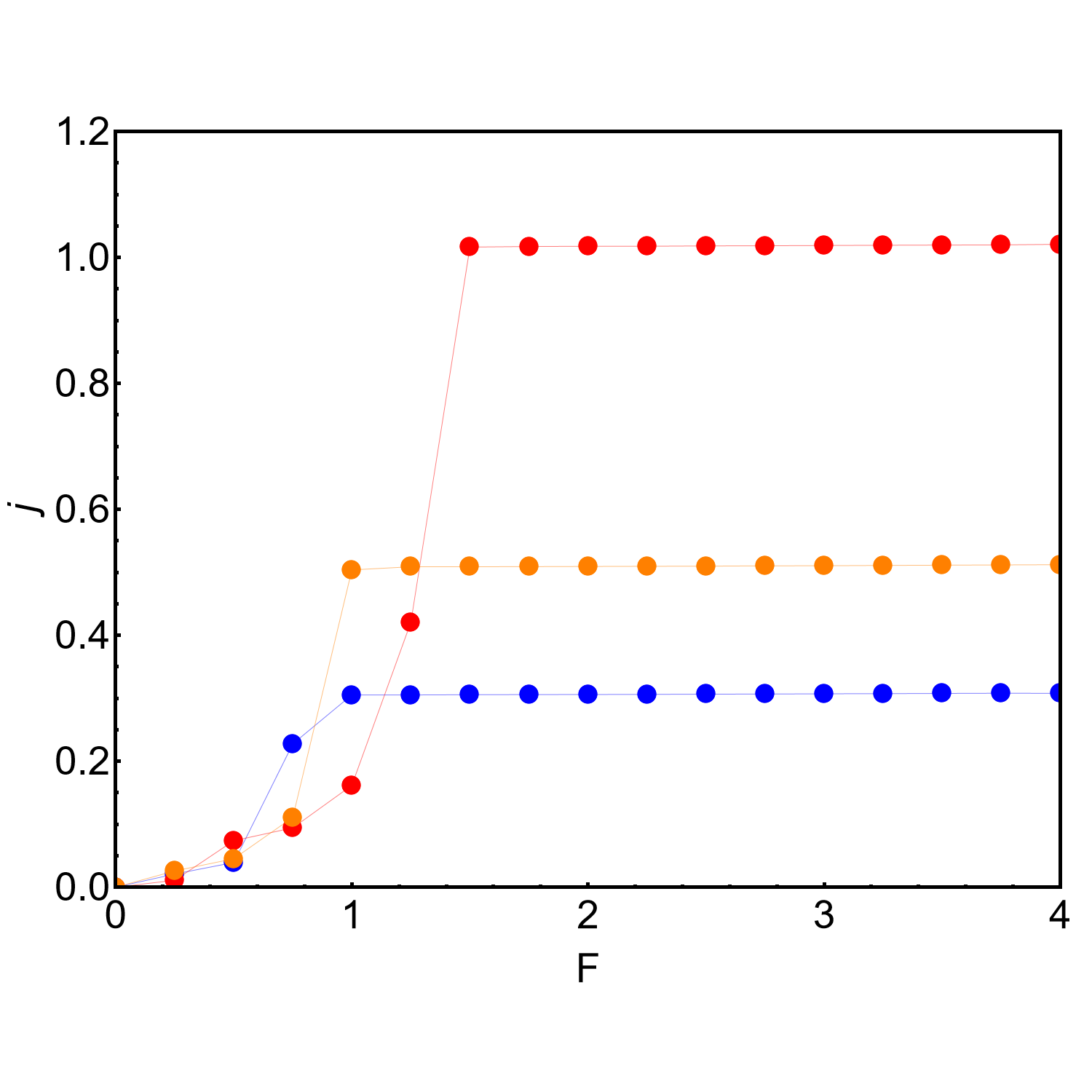} 
  \caption{$j$ vs. $F$. Left: bounded potential, $V(q)=q^4/(4(1+q^4))$, and  $\nu=-0.6$ (black dots), $-0.3$ (blue dots), $0.3$ (orange dots), $0.6$ (pink dots) and $1$ (red dots). Right:   unbounded potential, $V(q)=q^4/4$, and $\omega=0.3$ (blue dots), $0.6$ (orange dots) and $1$  (red dots).
 } \label{comj}        
 \end{figure}

The effect of $\nu>0$ in the overall behavior for both potentials
resembles qualitatively  to the aforementioned case $\nu=1$  (see
Figure \ref{comj}). The only difference appears when $\nu<0$,
where the only  unbounded case has a stationary state for small enough
values of $F$. In contrast, the bounded potential has well defined
stationary state for any value of $F$.
\section{Conclusions}
We have studied a finite anharmonic pinned chain of interacting
oscillators subjected to a periodic external forcing placed at the center of the lattice, with friction applied at both ends. 
In Section \ref{sec2}
some  recent rigorous results obtained in \cite{our}  are presented. Among others we
demonstrate a meaningful perturbative scheme for the anharmonic
problem that can be constructed around the exact harmonic solution,
provided two key conditions are met.
First, all the integer multiplicities of   the external forcing
  frequency must lie outside the spectrum of the respective infinite
harmonic chain (the interval made of frequencies corresponding to the
normal modes of the infinite harmonic pinned lattice). Second, both
the pinning and interaction potentials must have bounded second
derivatives. Under these conditions, it can be shown that 
the lower bound of the radius of convergence, $\nu_0$, is independent 
of the number of oscillators, the damping coefficient and the
magnitude of the forcing. As a concrete application, we implemented
the scheme to the single oscillator case, confirming the robustness of the perturbative approach. Additionally, we show that even this simple system exhibits a variety of complex behaviors when $|\nu|$ exceeds $\nu_0$. 

In the second part of the paper we present computer simulations
that explore the system behavior in regimes not covered by the
rigorous results. For instance, in the cases studied, when some
multiplicities of the force frequency lie  within the harmonic
spectrum interval, we observe that each mode of the stationary
periodic solution  behaves qualitatively similarly to its counterpart
in the purely harmonic case. Specifically, if a mode lies outside the
harmonic spectrum  interval, its amplitude decays exponentially
with the distance from the origin where the forcing is applied. In
contrast, if the mode lies within the harmonic spectrum interval, a plane-wave behavior emerges. 

We also have studied how the work depends on the forcing frequency and on the sign and magnitude  of the anharmonicity. In particular, we observe the phenomenon of supra-transmission for 
sufficiently large negative values of $\nu$. 

Finally, we emphasize the importance of distinguishing, in general,
between anharmonic potentials with bounded or unbounded second
derivatives. This distinction has played a crucial role in our
rigorous proof, see \cite{our}. In particular, we compare our findings with the numerical results of Prem et al. \cite{Prem} who simulated a system with unbounded pinning potential,  $V(q)=q^4/4$, and a forcing frequency within the harmonic interval. They observed that the heat current  exhibits a distinctive behavior as a function of the forcing magnitude,$F$: for $F\in [0,F_{c,1}]$ the current increases as $F^2$;  for $F\in [F_{c_1},F_{c,2}]$ it remains constant and for $F>F_{c,2}$ it decreases. In contrast, we find that this behavior disappears when performing the same simulation with a bounded potential: in that case, the heat current increases monotonically with $F$. This result suggests that the boundedness on the second derivative of the anharmonic potential may not merely be a technical requirement for a rigorous proof, but it could also have significant physical implications.

\section{Acknowledgments}
J.L.L thanks D. Huse and A. Dhar for useful discussions.
P.G.  acknowledges the support of 
  the Project I+D+i Ref.No. PID2023-149365NB-I00, funded by
  MICIU/AEI/10.13039/501100011033/ and by ERDF/EU, T.K  acknowledges the support of the NCN grant 2024/53/B/ST1/00286

  \bibliographystyle{amsalpha}

\section{Appendix: The harmonic solution}
Let
$$
\tilde q_{x,p}(m)=\frac{1}{\theta}\int_{0}^\theta dt e^{-im\omega
  t}q_{x,p}(t)
$$
be the time harmonics of the periodic solution, $\tilde q_{x,p}(m)=\tilde q_{x,p}(-m)^*$. It can be shown that for the harmonic case, $\nu=0$,
$\tilde q_{x,p}(m)$ is proportional to the forcing modes. In particular, for $F(t)=F\cos(\omega t)$ and $N\rightarrow\infty$
 the $\theta$-periodic solution has only two modes, $m\pm1$:
 \begin{eqnarray}
  \tilde q_x(1)=&=&\frac{F}{2D(\omega)}\xi_+^{-\vert x\vert}\quad, \omega<\omega_0\nonumber\\
 &=&\frac{F}{2iD(\omega)}e^{-i \phi(\omega)\vert x\vert}\quad, \omega\in\mathcal{I}=[\omega_0,\sqrt{\omega_0+4}]\nonumber\\
 &=&-\frac{F}{2D(\omega)}\xi_-^{-\vert x\vert}\quad, \omega>\sqrt{\omega_0+4}
 \end{eqnarray}
 where
 \begin{equation}
 D(\omega)=\left[\vert(\omega_0^2-\omega^2)(\omega_0^2+4-\omega^2)\vert\right]^{1/2}
 \end{equation}
 \begin{equation}
 \xi_\pm=1+\frac{1}{2}(\omega_0^2-\omega^2)\pm\frac{D(\omega)}{2}
 \end{equation}
 and
 \begin{equation}
 \omega^2=\omega_0^2+4\sin^2\left(\frac{\phi(\omega)}{2}\right)
 \end{equation}

\end{document}